\documentclass [prl,twocolumn,notitlepage,superscriptaddress,floatfix] {revtex4-1}
\usepackage{amsmath}
\usepackage{graphicx}
\usepackage{lmodern}
\usepackage{amsmath}

\usepackage{color}
\usepackage{amssymb}
\newcommand{\beq} {\begin{equation}}
\newcommand{\eeq} {\end{equation}}
\newcommand{\bea} {\begin{eqnarray}}
\newcommand{\eea} {\end{eqnarray}}
\newcommand{\be} {\begin{equation}}
\newcommand{\ee} {\end{equation}}
\renewcommand{\(}{\left(}
\renewcommand{\)}{\right)}
\renewcommand{\[}{\left[}
\renewcommand{\]}{\right]}

\DeclareMathOperator{\Tr}{Tr}
\DeclareMathOperator{\Ree}{Re}
\DeclareMathOperator{\Imm}{Im}
\newcommand{\su} {{\mathrm{SU(2)}}}
\newcommand{\so} {{\mathrm{SO(4)}}}

\begin{document}

\title {Coexistence of charge-density-wave and pair-density-wave orders in underdoped cuprates}
\author{Yuxuan Wang}
\affiliation{Department of Physics, University of Wisconsin, Madison, WI 53706, USA}
\author{Daniel F. Agterberg}
\affiliation{Department of Physics, University of Wisconsin-Milwaukee, Milwaukee, Wisconsin 53211, USA}
\author{Andrey Chubukov}
\affiliation{William I. Fine Theoretical Physics Institute,
and School of Physics and Astronomy,
University of Minnesota, Minneapolis, MN 55455, USA}
\date{\today}

\begin{abstract}
We analyze incommensurate charge-density-wave (CDW) and pair-density-wave (PDW) orders with transferred momenta $(\pm Q,0)$/$(0,\pm Q)$
 in underdoped cuprates within the
  spin-fermion model. Both orders appear due to exchange of spin fluctuations before magnetic order
  develops.
  We argue that
   the ordered state with the lowest energy has non-zero CDW and PDW  components with the same momentum.
Such a state breaks $C_4$ lattice rotational symmetry, time-reversal symmetry, and mirror symmetries. We argue that the feedback from CDW/PDW order on fermionic dispersion  is consistent with ARPES data.
 We discuss the interplay between the CDW/PDW order and $d_{x^2-y^2}$ superconductivity and make specific predictions for experiments.
\end{abstract}
\maketitle

{\it Introduction.}~~~ The search for
competitors to d$_{x^2-y^2}$ superconductivity (d-SC) in underdoped cuprates has gained strength over the last few years due to mounting experimental evidence that
 some form of  electronic charge order  spontaneously emerges below a certain doping and competes with d-SC
  (Refs.\ \cite{mark_last,ybco,ybco_1,X-ray, X-ray_1,davis_1,tranquada,ber09,shen_a,shen_2010,kerr,bourges,armitage,ando,hinkov,taillefer_last})
 The
  two most frequently discussed candidates for electronic order
   are incommensurate
 charge density-wave (CDW) order
 (Refs.\ \cite{grilli,ddw,ms,efetov,greco,laplaca,charge,tsvelik,norman,debanjan,pepin,charge_1,atkinson,rahul,debanjan_1,pepin_new})
  and
  incommensurate pair-density-wave order
  (PDW), which is a SC order with a finite Cooper pair momentum
  ${\bf Q}$ (Refs.~\cite{kivelson,agterberg,agterberg_2,patrick,lee_senthil,corboz}).
   Other potential  candidates
    are loop current order~\cite{varma} and CDW order with momentum near $(\pi,\pi)$ (Ref. \cite{sudip}).
    
CDW order in underdoped cuprates has been proposed some time ago~\cite{grilli} and has been analyzed in detail by several groups in the last few years within the spin-fluctuation formalism~\cite{ms,efetov,charge,tsvelik,laplaca,debanjan,pepin,charge_1}
and within $t-J$ model~\cite{ddw,greco}.
  The initial discussion was focused on near-equivalence between d-SC and
 d-wave charge bond order (BO)
with  momenta $(Q,Q)$ along zone diagonal~\cite{ms,efetov,pepin}, but charge order of this type has not been observed in the experiments.
It was later found~\cite{laplaca,debanjan,charge,charge_1} that the same magnetic model also displays a
 CDW order with momenta $(Q,0)$  or $(0,Q)$, which is consistent with the range of CDW wave vectors extracted from experiments~\cite{mark_last,ybco,ybco_1,X-ray, X-ray_1,davis_1,shen_a, shen_2010,proust}.
   Such CDW order is also consistent with experiments that detect the breaking of discrete rotational and time-reversal symmetries in a $(T,x)$ range where competing order develops \cite{kerr,bourges,armitage,ando,hinkov,taillefer_last}.  In particular, when spin-fermion coupling is strong enough,
     the CDW order develops in the form of a stripe and  breaks $C_4$  lattice rotational symmetry.
     A stripe CDW order with $(Q,0)/(0,Q)$ in turn
     gives rise to modulations in both charge density and charge current and
      breaks time-reversal and mirror symmetries~\cite{charge,charge_1,tsvelik,rahul}.

  The agreement with the data is encouraging, but two fundamental issues with CDW order remain.  First, within the mean-field approximation,
  $T_{\rm cdw}$ is smaller than the superconducting $T_c$ (and also the onset temperature for $(Q,Q)$ order.
 It has been conjectured that $T_{\rm cdw}$ may be enhanced  by adding e.g., phonons~\cite{grilli}, or nearest-neighbor Coulomb interaction~\cite{allias} or
 assuming the CDW emerges from already pre-existing  pseudogap~\cite{atkinson,debanjan}. $T_{\rm cdw}$ is also enhanced  by fluctuations beyond mean-field~\cite{charge,tsvelik},
  but whether such enhancements are strong
   enough to make $T_{\rm cdw}$ larger than $T_c$ remains to be seen.
     Second, stripe CDW order
      cannot
       explain
       qualitative features of the ARPES data
        away from zone boundaries~\cite{patrick}.

It has been argued~\cite{patrick} that ARPES experiments for all momentum cuts can be explained by assuming that the competing order is PDW rather than CDW.
PDW order was initially analyzed for doped Mott insulators~\cite{kivelson,lee_senthil,corboz}, but it also emerges in the spin-fermion model~\cite{charge_1}  with the same momentum $(Q,0)/(0,Q)$ as CDW order and its onset temperature $T_{\rm pdw}$ is close to $T_{\rm cdw}$ (the two become equivalent if one neglects the curvature of fermionic dispersion at hot spots~\cite{pepin,charge_1}).
   Given that PDW order explains ARPES experiments, it seems logical to consider it as a candidate for competing order.
  Just like CDW, the PDW order
    develops in the form of a stripe and breaks $C_4$ lattice rotational symmetry~\cite{agterberg,charge_1}, if, again, the coupling is strong enough.
     However, it does not naturally break time-reversal and mirror symmetries~\cite{agterberg_2}
 (although it does so for a particular Fermi surface geometry~\cite{agterberg}), and  the mean-field $T_{\rm pdw}$ is also smaller than $T_c$ for d-SC.
     \begin{figure}
  \includegraphics[width=\columnwidth]{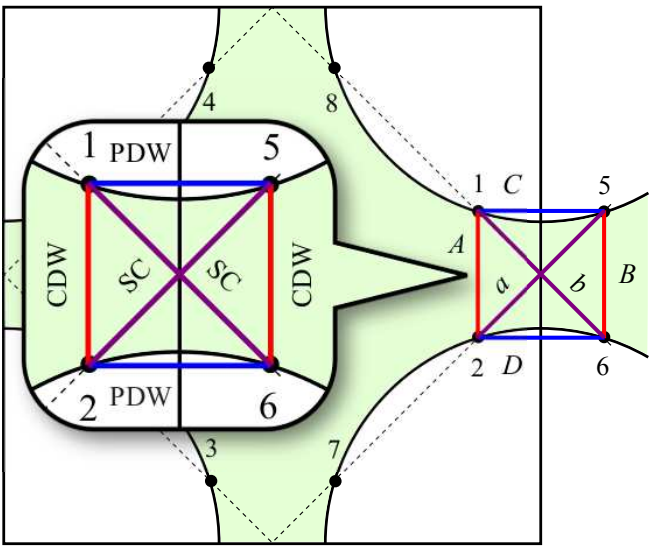}
     \caption{
     The Brillouin zone, the Fermi surface, and the hot spots.
   We label bonds connecting hot spots as $A$, $B$, $C$, $D$, $a$ and $b$. Inset: the structure of the mixed CDW/PDW state
     in one of the hot regions.}
     \end{figure}

In this communication we build on the results of the generic Ginzburg-Landau analysis
~\cite{charge_1} and
 propose how to resolve the partial disagreement with experiments for pure CDW or
PDW orders.
 We first re-iterate that pure CDW/PDW
  orders emerge in the forms of stripes only if the spin-fermion interaction $g$ is strong enough. In practice,
   $g$
  has to be at least comparable to the upper energy cutoff of the spin-fermion model $\Lambda$ (see details below). For smaller couplings
 the system  develops a checkerboard order
   for which $C_4$ symmetry is preserved~\cite{comm}.    The spin-fermion model is a low-energy model and it is rigorously defined only when the coupling $g$ is smaller than $\Lambda$.  In this respect,  stripe CDW or PDW orders emerge,
 only at the
  edge
  of the applicability of the model.
 Here we consider spin-fermion model at smaller couplings, well within its applicability range, and allow both CDW and PDW orders to develop.
  We show that the system
  develops a
     mixed CDW/PDW
     order, in which a CDW component develops between hot fermions separated along, say,
      Y
      direction and a PDW component develops between fermions separated along
       X
       direction (see Fig. 1). Because the momentum  carried by an order parameter is the transferred momentum for CDW and the total momentum for PDW, the CDW order along
       Y and the PDW order along
       X
        actually carry the same momentum
        $(0,Q)$.
         We argue that such a state  further
  lowers its Free energy by developing (via an emerging triple coupling) secondary homogeneous superconducting orders~\cite{charge_1}.
  This effect favors the mixed CDW/PDW state over
      the pure checkerboard CDW or PDW states, which would otherwise all be degenerate.
        The mixed CDW/PDW state breaks $C_4$ symmetry because both orders carry either momentum $(Q,0)$ or $(0,Q)$, but not both, and it also
         breaks time-reversal and mirror symmetries as the pure stripe CDW order with $(Q,0)$ or $(0,Q)$ does.

    The presence of PDW component is
    relevant for the interpretation of the ARPES data. Without it, the fermionic spectrum in the CDW phase would contain
    the lower energy  branch,
    which never crosses Fermi level, and the  upper energy branch, which would approach the Fermi level {\it from above} as the momentum cuts enter the arc region.
     As discussed in~\cite{patrick},  this is inconsistent with the data~\cite{shen_a}  which show that  the dispersion
      approaches the Fermi level {\it from below}.
      We show that the presence of PDW component changes the structure of fermionic dispersion in such a way that now the lower branch crosses the Fermi level in the arc region (see Fig.\ 2), in full agreement with ARPES experiments.

  We also consider the interplay between CDW/PDW order and d-SC and present the
   phase diagram in Fig.\ 3.
    The reduction of the superconducting $T_c$ in the coexistence region with CDW/PDW is the obvious consequence
     of competition for the Fermi surface. A small (of order $g/\Lambda$) drop of
       $T_c$ upon entering the coexistence region is the result of a weak first-order CDW/PDW transition.
       There
        exists,
        however, a more subtle feature of the phase diagram.
        Namely,
        a secondary SC  order
         is generated by CDW/PDW order,
     which preserves the same sign of the gap along each quadrant of the Fermi surface.
      Below $T_c$ for d-SC, this secondary superconducting order couples with  $d_{x^2 -y^2}$ order, and the net result is the removal or shifting of the gap nodes.
      Simultaneously, the CDW order acquires an extra component with $s$-form factor, i.e., the magnitude of its s-wave portion increases.  
 We propose to verify these through experiments.

 {\it The model}~~~~We follow previous works~\cite{ms,efetov,charge,charge_1} and consider emerging charge order within the spin-fermion model~\cite{acs}. This model describes interactions between itinerant electrons and their near-critical antiferromagnetic collective spin excitations in two spatial dimensions.
  Eight ``hot" spots, defined as points on the Fermi surface separated by antiferromagnetic ordering momentum $(\pi,\pi)$ (points 1-8 in Fig.\ 1), are the most
   relevant for  destruction of a normal Fermi liquid state.  The known instabilities of the spin-fermion model include d-SC (e.g.\ $\langle c_1 c_6\rangle$,
    see Fig.\ 1)~\cite{ms, wang,wang_el}, bond charge order (BO) with momenta $(\pm Q, \pm Q)$ (e.g.\ $\langle c_1^\dagger c_6\rangle$)~\cite{ms, efetov,pepin},
    CDW order with momenta $(0,\pm Q)$ and $(\pm Q,0)$ (e.g.\ $\langle c_1^\dagger c_2\rangle$)~\cite{charge,debanjan,debanjan_1} and PDW order
     with momenta $(0,\pm Q)$ and $(\pm Q,0)$ (e.g.\ $\langle c_1 c_2\rangle$)~\cite{pepin,charge_1}.
        The model has an approximate $\su$ particle-hole symmetry~\cite{ms,efetov,pepin,charge_1,pepin_new}, which becomes exact once one linearizes the fermionic dispersion in the vicinity of the hot spots. This gives rise to near-degeneracy between d-SC and BO and between CDW and PDW.

 {\it The Ginzburg-Landau analysis}~~~~We introduce four order parameters: $\Psi$ for SC, $\Phi$ for BO, $\psi$ for PDW, and $\rho$ for CDW respectively. SC and BO order parameters connects hot spots along diagonal bonds, which we label as $a$ and $b$, while PDW and CDW connect hot spots along vertical and horizontal bonds, which we label as $A$-$D$ in Fig. 1. We define the CDW order parameter on bond $A$ as
 $\rho_A\sim \langle c_1^\dagger c_2 \rangle$ and use analogous notations for other order parameters.
The effective action has three terms:
\begin{align}
{\mathcal S}_{\rm eff}=&{\mathcal S}_{\rm cdw/pdw} [\rho,\psi]+{\mathcal S}_{\rm sc/bo}[\Psi,\Phi]
+{\mathcal S}_{\rm int}
\label{seff_1}
\end{align}
\begin{figure}
\includegraphics[width=\columnwidth]{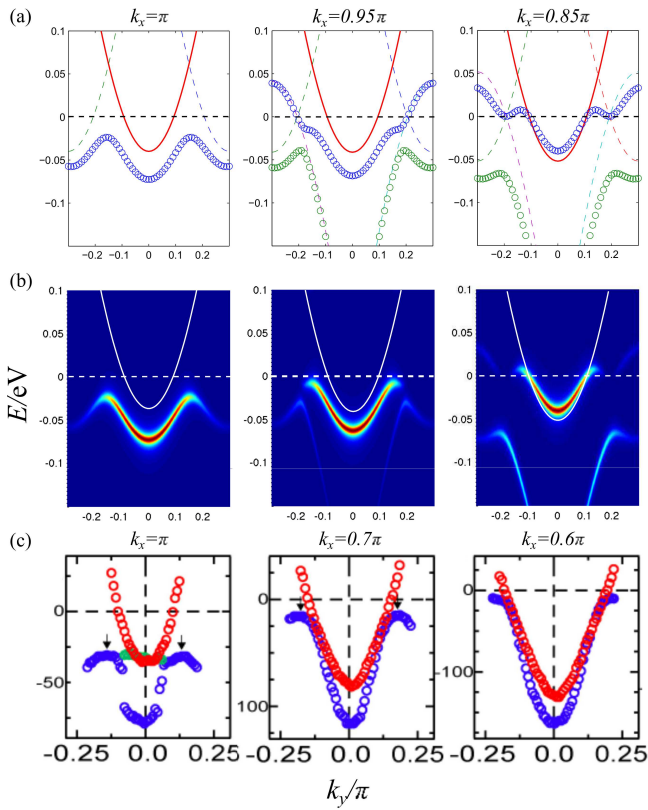}
\caption{Fermionic dispersion in the antinodal region in the presence of the mixed CDW/PDW order. Upper panel -- the dispersion in the presence of CDW/PDW order
 for various $k_x$ ($k_x =\pi$ corresponds to the cut along the Brillouin zone boundary). Middle panel-- the spectral function. Thin line on both panels is the bare dispersion. Bottom panel -- experimental data from Ref.\ \cite{shen_a} for comparison.  The experimental data have been taken below $T_c$ and show a gapped dispersion in a wider range of $\pi - k_x$. }
\end{figure}
The ${\mathcal S}_{\rm cdw/pdw} [\rho,\psi]$ term is of our primary interest. Keeping the SU(2) symmetry exact, we follow Ref. \cite{charge_1} and
combine  PDW and CDW orders on a given bond (say, bond A) into a $2\times2$ matrix order parameter
     \begin{align}
&\Delta_A^{\mu\nu}\equiv\(\begin{array}{cc}
\psi_A&\rho_A^*\\
-\rho_A&\psi_A^*
\end{array}\)\equiv \sqrt{|\rho_A|^2+|\psi_A|^2}~U_A,
\end{align}
where $\rho_A\sim c_1^\dagger c_2$, $\psi_A\sim c_1c_2$, and $U_A$ is a $\su$ matrix ``phase".
 The order parameters $\Delta_{B,C,D}$ and phases $U_{B,C,D}$ are similarly defined~
 (see Supplementary Material (SM) for details).
  Minimizing the Free energy, we obtain $\Gamma\equiv\Tr(U_AU_C^\dagger U_BU_D)=-2$, $\sqrt{|\rho_A|^2+|\psi_A|^2}=\sqrt{|\rho_B|^2+|\psi_B|^2}\equiv|\Delta_y|$, and $\sqrt{|\rho_C|^2+|\psi_C|^2}=\sqrt{|\rho_D|^2+|\psi_D|^2}\equiv|\Delta_x|$. Under these conditions, the CDW/PDW action becomes
\begin{align}
\mathcal{S}_{\rm cdw/pdw}=&\frac{\alpha}{2} (|\Delta_x|^2+|\Delta_y|^2)+\beta(|\Delta_x|^4+|\Delta_y|^4)\nonumber\\
&+(\tilde\beta-\bar\beta)|\Delta_x|^2|\Delta_y|^2+O(\Delta^6)
\label{seff}
\end{align}
where $\alpha\sim\Lambda/{v_F^2}\times(T-T_{\rm cdw})/T_{\rm cdw}$ and $T_{\rm cdw} = T_{\rm pdw} \sim g$ (Ref. \cite{charge}).
The prefactors $\beta$, ${\tilde \beta}$, and ${\bar \beta}$  are determined by different convolutions of four fermionic propagators (the square diagrams ~\cite{charge,debanjan_1,charge_1}).  At $g \ll \Lambda$ we have $\beta \sim 1/(v_F^2\Lambda)$, ${\tilde \beta} \sim \log(\Lambda/g)/(v_F^2\Lambda)$, and
${\bar\beta} \sim (\Lambda/g)/(v_F^2 \Lambda)$.  We see that ${\bar \beta}$ is the largest term, hence the action (\ref{seff}) is minimized when
 $|\Delta|\equiv|\Delta_x|=|\Delta_y|$. Because $\tilde\beta-\bar\beta <0$, the action is unbounded, which implies that the transition is first-order and sixth-order terms (coming from six-leg diagrams) have to be included  to stabilize the order.  Including these terms we obtain
 a first order into CDW/PDW state at $T_{\rm cdw/pdw} = T_{\rm cdw} (1 + O(g/\Lambda))$.
   We emphasize that this temperature is higher than the one for a pure CDW (or PDW) transition.

The constraint $\Gamma\equiv\Tr(U_AU_C^\dagger U_BU_D)=-2$ leaves the ground state hugely degenerate -- the order parameter manifold is $\so\times\so$ (Ref.\ \onlinecite{charge_1}).  This manifold includes pure CDW and pure PDW checkerboard states and mixed CDW/PDW states.  To select the actual ground state configuration we note that, if CDW and PDW orders have components which carry the same momentum ${\bf Q}$,  the Free energy is further
lowered by creating a secondary order whose magnitude is a product of CDW and PDW order parameters.
 This secondary order is a homogeneous SC with equal sign of the gap along each quadrant of the FS~\cite{charge_1}
  One can straightforwardly check that
   the reduction of the Free energy is maximal when
    in a nominally checkerboard state CDW occurs along vertical bonds and PDW occurs along horizontal bonds or vise versa, i.e., each order develops in the form of a stripe.
  This corresponds to either $\psi_{A,B}=\rho_{C,D}=0$ (as in the inset of Fig. 1) or
  $\psi_{C,D}=\rho_{A,B}=0$, the choice breaks $C_4$ lattice rotation symmetry.
  Furthermore, the stripe CDW order parameters $\rho_A$ and $\rho_B$ and PDW order parameters $\psi_C$ and $\psi_D$ get separately coupled by
  fermions away from hot spots, and the coupling between $\rho_A$ and $\rho_B$  locks the relative phase of $\rho_A$ and $\rho_B$ such that $\rho_B = \pm i \rho_A$ (Ref. \cite{charge}). The choice of the sign breaks time-reversal and mirror symmetries.  The coupling between $\psi_{C}$ and $\psi_D$ does not lock  their phases.

\begin{figure}
\includegraphics[width=\columnwidth]{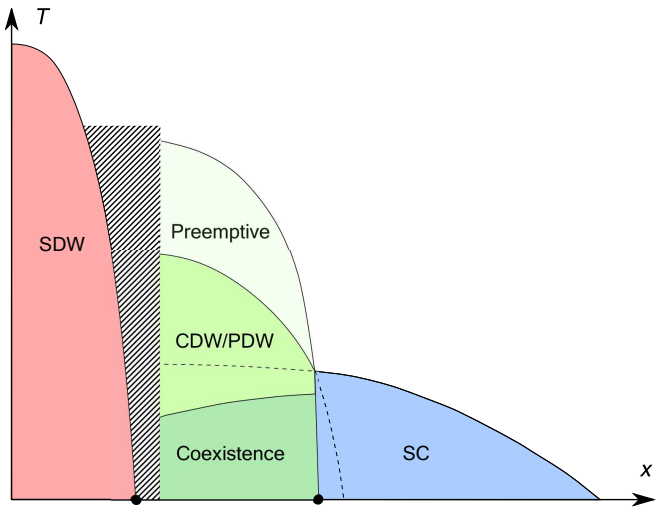}
 \caption{The phase diagram.  The transition into CDW/PDW state is weakly first-order and the superconducting $T_c$  drops by a finite amount
   upon entering into coexistence region.  In the region labeled as ``pre-emptive"
  discrete $C_4$ and time-reversal/mirror symmetries are broken but continuous $U(1)$ translational symmetry (associated with the locking of the common phases
   of $\rho_A$ and $\rho_B$) remains unbroken~\cite{charge}.
    In the shaded region, Mott physics
    develops
    and the
     onset temperature of charge ordering shrinks.
   }
\end{figure}

{\it Feedback from CDW/PDW order on fermions}~~~~We now show that the feedback from stripe CDW/stripe PDW order on the
fermionic dispersion
 at $k\sim(\pi,0)$, taken as a function of $k_y$ for various $k_x=\pi-\delta k_x$, yields results
  in quite reasonable agreement
  with ARPES data~\cite{shen_a,shen_2010}.
 Previous studies have shown~\cite{charge} that a pure CDW order can explain the ARPES spectrum for a cut along the BZ boundary, but not for cuts that are closer toward BZ center
 (see Ref.~\cite{patrick,sm}).
To obtain the dispersion along various cuts in the presence of both CDW and PDW, we have extended our analysis of the CDW/PDW order
 to a finite momentum range
 away from the hot spots.
We find that at the BZ boundary, the CDW order has a  larger amplitude  due to better FS nesting but
  the PDW component increases as the cuts move towards the hot spots.
We present the details in SM
 and show the results in Fig. 2.
  There are three key features in our scenario that are
  qualitatively
  consistent with experiment: (1) at the BZ boundary ($k_x=\pi$), the locus of minimum excitation energy shifts from $k_F$ to a larger value $k_G\approx Q/2$, where $Q$ is the CDW momentum, (2) as $k_x$ decreases, the excitation approaches the  Fermi level from below,  and (3) at $k_x$ when the Fermi arc emerges, the fermionic dispersion becomes flat for $|k_y|>k_F$.
   These features are also reproduced by pure PDW order \cite{patrick} and from a spatially homogeneous self-energy arising from a $d$-wave CDW order peaked at $(\pi,\pi)$ \cite{greco}. However, both these scenarios
  do not immediately explain
  the observation of broken time-reversal symmetry or
   CDW order with small incommensurate momentum.
    To obtain quantitative agreement with the experiments, we would need to know how CDW and PDW order parameters depend on frequency. This would require one to model the bare dispersion far away from $k_F$ and solve complex integral equations for frequency-dependent order parameters.

{\it Interplay between CDW/PDW order and $d_{x^2 -y^2}$ superconductivity}~~~~
 We next consider other terms in the effective action in Eq. (\ref{seff_1}).  The term $\mathcal{S}_{\rm sc/bo}$ has been analyzed in ~\cite{efetov,pepin,debanjan_1}.
 When $\su$ symmetry is exact, d-SC and BO orders are degenerate and the action has
  four Goldstone modes.
  Once $\su$ symmetry is broken by  FS curvature, only d-SC order  develops below $T_c$. We assume that this is the case and keep only
   d-SC component $\Psi$ in $\mathcal{S}_{\rm sc/bo}$, i.e. reduce it to $\mathcal{S}_{\rm sc/bo} = \alpha_s|\Psi|^2+ \beta_s|\Psi|^4$ with $\alpha_s \sim\Lambda/v_F^2\times(T-T_c)/T_c$, $T_c \sim g$, and $\beta_s\sim \Lambda/(v_F g)^2$.  The coupling between CDW/PDW and d-SC orders is again obtained by  evaluating
    the square diagrams. The calculation yields
     $\mathcal{S}_{\rm int} = \beta' |\Delta|^2|\Psi|^2$ with $\beta' \sim1/(v_F^2 g)$.
      Note that the magnitude of the coupling is phase sensitive, hence the phase locking between $\rho_A$ and $\rho_B$
     at $\pm \pi/2$
      is important (see SM for details).

  The analysis of the full action
  is straightforward and we show the results in Fig.\ 3.
  The mean-field temperature $T_{\rm cdw/pdw} \geq T_{\rm cdw}$ is comparable to $T_c$ near the SDW boundary but is
    enhanced by fluctuations~\cite{debanjan,atkinson,charge}.  We assume that
  this enhancement
   lifts
  $T_{\rm cdw/pdw}$ above $T_c$ at large $\xi$.
 Because CDW/PDW transition is first-order, $T_c$ jumps
 upon entering into the coexistence region, but the jump is again small in $g/\Lambda$.
 Similar behavior has been recently observed in Fe-pnictides~\cite{pnictides}.
 At small $T$, the CDW/PDW and d-SC orders coexist.

The phase diagram in Fig.\ 3 is similar to that for pure CDW order~\cite{charge}, but there are some extra features. 
 First,
the combination of CDW/PDW orders induces a secondary SC order~\cite{charge_1} with a non-zero gap along zone diagonal ($s$-wave or $d_{xy}$).
 In the coexistence region %with d-SC 
 the
   order $\Psi_s$ couples with d-SC order $\Psi$ and, as a result, gap nodes either get shifted ($d+s$ state) or removed ($d + e^{i\theta} s$ state).
A similar coupling has been examined in the context of the Fe-pnictides \cite{hinojosa}. 
 A finite gap along zone diagonals has been observed in ARPES at doping $x<0.1$ (Ref. \cite{shen_b}) and also inferred from Raman spectroscopy \cite{sakai}.
    Second, by the same logic, the d-SC and PDW orders induce a secondary $s$-wave CDW order with the same momentum as the primary one. We propose to search for SC gap opening or node shifting and to examine the $s$-component of CDW order in the coexistence region.

{\it Conclusions}~~~~In this letter we proposed a state with unidirectional CDW and PDW orders which carry the same momentum.
  We argued that this state is a member of the ground state manifold of the low-energy spin-fermion model and its energy is further reduced by induction of a
   secondary SC order. We further argued that CDW/PDW state has a number of features  consistent with experiments: it breaks both $C_4$ and time-reversal symmetry and the feedback from CDW/PDW order on fermions reproduces the ARPES data from the BZ boundary to the tip of the Fermi arc. The transition into CDW/PDW state is weakly first-order and occurs at a higher transition temperature than that for a pure unidirectional CDW or PDW orders. We considered the interplay between CDW/PDW order and d-SC, and found that a SC gap becomes non-zero along zone diagonals. We proposed to search for this gap opening
   in the region where charge order and d-SC coexist.

  We acknowledge useful discussions with W.\ A.\ Atkinson, D.\ Chowdhury, E. \ Fradkin,  S.\ Kivelson, P.\ A.\ Lee,  C.\ P\'epin, and S.\ Sachdev, The
  work was supported by the DOE grant DE-FG02-ER46900 (AC and YW) and by NSF grant No. DMR-1335215 (DFA).

\onecolumngrid

\newpage
\centerline{\large{\bf Supplementary Material}}

\section{I.~~~Details of the Ginzburg-Landau action}
\subsection{A.~~~CDW/PDW sector}
The Ginzburg-Landau (GL) action for the CDW/PDW order has been derived and studied in detail in Ref.~\onlinecite{sm_pdw}.
We  briefly review the analysis here and apply it for our purposes.

When the curvature of the Fermi surface (FS) at hot spots can be neglected, spin-fermion model has $\su$ symmetry which makes CDW and PDW orders degenerate.
 The action of the spin-fermion model can be rewritten in an explicitly $\su$-symmetric form in terms of particle-hole doublets at each of the hot spots 1-8,
\begin{align}
&\Psi_1({k})=\(\begin{array}{c}
c_{1\uparrow}({k})\\
c_{1\downarrow}^{\dagger}(-{k})
\end{array}\),~~
\Psi_2({k})=\(\begin{array}{c}
c_{2\downarrow}^\dagger(-{k})\\
c_{2\uparrow}({k})
\end{array}\)\nonumber\\
&\Psi_3({k})=\(\begin{array}{c}
c_{3\uparrow}({k})\\
c_{3\downarrow}^{\dagger}(-{k})
\end{array}\),~~
\Psi_4({k})=\(\begin{array}{c}
c_{4\downarrow}^\dagger(-{k})\\
c_{4\uparrow}({k})
\end{array}\)\nonumber\\
&\Psi_5({k})=\(\begin{array}{c}
c_{5\downarrow}^\dagger(-{k})\\
c_{5\uparrow}({k})
\end{array}\),~~
\Psi_6({k})=\(\begin{array}{c}
c_{6\uparrow}({k})\\
c_{6\downarrow}^{\dagger}(-{k})
\end{array}\)\nonumber\\
&\Psi_7({k})=\(\begin{array}{c}
c_{7\downarrow}^\dagger(-{k})\\
c_{7\uparrow}({k})
\end{array}\),~~
\Psi_8({k})=\(\begin{array}{c}
c_{8\uparrow}({k})\\
c_{8\downarrow}^{\dagger}(-{k})
\end{array}\).
\end{align}
In this notation, the CDW order parameter $\rho$'s and the PDW order parameter $\psi$'s involving the same pair of hot spots (labeled as $A$, $B$, $C$, and $D$ in Fig.\ 1 in the main text) can be combined into a 2$\times$2 matrix  that couples bilinearly to particle-hole doublet $\Psi$'. The four 2$\times$2 matrices are
\begin{align}
\Delta_A^{\mu\nu}=\(\begin{array}{cc}
\psi_A&\rho_A^*\\
-\rho_A&\psi_A^*
\end{array}\),~
\Delta_B^{\mu\nu}=\(\begin{array}{cc}
\psi_B&\rho_B\\
-\rho_B^*&\psi_B^*
\end{array}\),~
\Delta_C^{\mu\nu}=\(\begin{array}{cc}
\psi_C&\rho_C^*\\
-\rho_C&\psi_C^*
\end{array}\),~\Delta_D^{\mu\nu}=\(\begin{array}{cc}
\psi_D&\rho_D\\
-\rho_D^*&\psi^*_D
\end{array}\).
\end{align}
For convenience, we also define $\su$ phases $U_{A,B,C,D}$ via $U_{A,B,C,D}\equiv\Delta_{A,B,C,D}/\sqrt{|\rho_{A,B,C,D}|^2+|\psi_{A,B,C,D}|^2}$.
Each order parameter changes sign under a momentum shift of $(\pi,\pi)$ (e.g., between the pair 1,2 and the pair 3,4 in Fig. 1 in the main text) because spin-mediated interaction is repulsive.  The magnitudes of the CDW and PDW order parameters between 1,2 and 3,4 do not have to match as these
 two pairs of hot spots are not equivalent. For simplicity, below we neglect this non-equivalence and assume that order parameters just change sign under a momentum shift by $(\pi,\pi)$ (this is often termed the $d-$wave approximation for the form-factor of the charge order).

The full effective action in terms of  CDW and PDW order parameters  up to quartic order is
\begin{align}
\mathcal{S}_{\rm eff}=&\alpha' \Tr(\Delta_A^\dagger \Delta_A+\Delta_B^\dagger \Delta_B+\Delta_C^\dagger \Delta_C+\Delta_D^\dagger \Delta_D)\nonumber\\
&-(I_1+I_2)\Tr(\Delta_A\Delta_A^\dagger\Delta_A\Delta_A^\dagger+\Delta_B\Delta_B^\dagger\Delta_B\Delta_B^\dagger+\Delta_C\Delta_C^\dagger\Delta_C\Delta_C^\dagger+\Delta_D\Delta_D^\dagger\Delta_D\Delta_D^\dagger)\nonumber\\
&-2I_3\Tr\[(\Delta_A\Delta_A^\dagger+\Delta_B\Delta_B^\dagger)(\Delta_C\Delta_C^\dagger+\Delta_D\Delta_D^\dagger)\]-4I_4\Tr\(\Delta_A^\dagger\Delta_B\Delta_C^\dagger\Delta_D\)\nonumber\\
=&2\alpha' (|\rho_A|^2+|\psi_A|^2+|\rho_B|^2+|\psi_B|^2+|\rho_C|^2+|\psi_C|^2+|\rho_D|^2+|\psi_D|^2)\nonumber\\
&-2(I_1+I_2)\[(|\rho_A|^2+|\psi_A|^2)^2+(|\rho_B|^2+|\psi_B|^2)^2+(|\rho_C|^2+|\psi_C|^2)^2+(|\rho_D|^2+|\psi_D|^2)^2\]\nonumber\\
&-4I_3(|\rho_A|^2+|\psi_A|^2+|\rho_B|^2+|\psi_B|^2)(|\rho_C|^2+|\psi_C|^2+|\rho_D|^2+|\psi_D|^2)\nonumber\\
&-4I_4\sqrt{(|\rho_A|^2+|\psi_A|^2)(|\rho_B|^2+|\psi_B|^2)(|\rho_C|^2+|\psi_C|^2)(|\rho_D|^2+|\psi_D|^2)}\Tr\(U_A^\dagger U_CU_B^\dagger U_D\)
\label{cpdw}
\end{align}

In mean-field analysis the coefficient $\alpha'$ is proportional to $g^{-1}_{\rm eff}-\Pi(T)$, where $g_{\rm eff}$ is the effective four-fermion interaction and $\Pi$ is the polarization operator (see Ref.~\onlinecite{sm_charge} for details). The polarization bubble $\Pi(T)$ increases as temperature decreases, and $\alpha'=\alpha' (T)$ changes sign at the CDW instability temperature $T=T_{\rm cdw}=T_{\rm pdw}$. By dimensional argument $\Pi \sim \Lambda/v_F^2 f(T/T_{\rm cdw})$, where $\Lambda$ is the upper cutoff of the spin-fermion model, and $f(x)$ is a dimensionless function. Then  $\alpha'= \Lambda/v_F^2(f(1)-f(T/T_{\rm cdw}))=\Lambda/v_F^2[(T-T_{\rm cdw})/T_{\rm cdw}]f'(1)\sim \Lambda/v_F^2[(T-T_{\rm cdw})/T_{\rm cdw}]$.

 The calculation of $T_{\rm cdw} = T_{\rm pdw}$ requires more care.  If one neglects momentum and  frequency dependence of $g_{\rm eff}$,  one obtains that $T_{\rm cdw}$ is non-zero only if ${g_{\rm eff}}$ exceeds some critical value \cite{sm_charge, 23}.  This is a
    consequence of the velocities of hot fermions at ${\bf k}_F$  and ${\bf k}_F + {\bf Q}$ being generally not antiparallel.
    However, once one includes the fact that spin-fermion interaction $g_{\rm eff} = g \chi (q, \Omega)$  is mediated by a boson with near-divergent dynamical
    susceptibility $\chi (q, \omega)$, one obtains~\cite{sm_charge} that
      the threshold vanishes when the magnetic correlation length diverges. In this limit, the CDW/PDW instability occurs for arbitrary values of the
       spin-fermion coupling and $T_{\rm cdw}=T_{\rm pdw}\sim g$. The spin-fermion model is justified as a low-energy model when interactions do not take a fermion outside of the low-energy subset, which holds when $g\ll \Lambda$.

\begin{figure}
\includegraphics[width=.8\columnwidth]{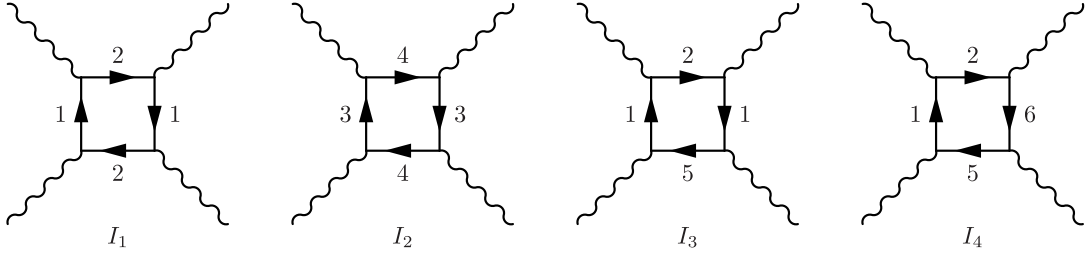}
\caption{The diagrams for coefficients $I_1$ - $I_4$.}
\label{I}
\end{figure}

The coefficients $I_{1,2,3,4}$ are obtained by evaluating the four square diagrams in Fig.\ \ref{I}. In explicit form,
\begin{align}
I_1=&-T\sum_{\omega_m}\int \frac{d^2k}{4\pi^2} G_1^2(\omega_m, k)G_2^2(\omega_m, k),\nonumber\\
I_2=&-T\sum_{\omega_m}\int \frac{d^2k}{4\pi^2} G_3^2(\omega_m, k)G_4^2(\omega_m, k),\nonumber\\
I_3=&-T\sum_{\omega_m}\int \frac{d^2k}{4\pi^2} G_1^2(\omega_m, k)G_2(\omega_m, k)G_5(\omega_m, k),\nonumber\\
I_4=&-T\sum_{\omega_m}\int \frac{d^2k}{4\pi^2} G_1(\omega_m, k)G_2(\omega_m, k)G_5(\omega_m, k)G_6(\omega_m, k),
\end{align}
where $G_i(\omega_m,k)$ is the Green functions for a fermion near hot spot $i$, and momentum $k$ is defined as a deviation from this hot spot.
The evaluation of the integrals yields~\cite{sm_charge, 23}
\begin{align}
I_1&=-\frac{1}{16\pi^2v_x^2v_y}\frac{1}{\Lambda_k},\nonumber\\
I_2&=0,\nonumber\\
I_3&=-\frac{1}{16\pi^2v_x^2v_y}\frac{1}{\Lambda_k}\log\frac{\Lambda}{T}\nonumber\\
I_4&=-\frac{1}{32v_xv_y}\frac{1}{T}.
\end{align}
where $v_x$ and $v_y$ are $x$ and $y$ components of the fermi velocity ${\bf v}_F$ at hot spot 1, and $\Lambda_k$ is the momentum cutoff $\sim \Lambda /v_F$.  Using the fact that $T\sim g$, we have $I_1\sim1/(v_F^2\Lambda)$, $I_3\sim1/(v_F^2\Lambda)\log(\Lambda/g)$ and $I_4\sim1/(v_F^2g)$.

 Because $U_A^\dagger U_CU_B^\dagger U_D\in \su$, the term proportional to $I_4$ in Eq.\ (\ref{cpdw}) is minimized when $\Tr\(U_A^\dagger U_CU_B^\dagger U_D\)=-2$.
 We also find that in the ground state $\sqrt{|\rho_A|^2+|\psi_A|^2}=\sqrt{|\rho_B|^2+|\psi_B|^2}\equiv|\Delta_y|$ and $\sqrt{|\rho_C|^2+|\psi_C|^2}=\sqrt{|\rho_D|^2+|\psi_D|^2}\equiv|\Delta_x|$. Under these conditions, Eq.\ (\ref{cpdw}) becomes
\begin{align}
\mathcal{S}_{\rm cdw/pdw}=\frac{\alpha}{2}(|\Delta_x|^2+|\Delta_y|^2)+\beta(|\Delta_x|^4+|\Delta_y|^4)+(\tilde\beta-\bar\beta)|\Delta_x|^2|\Delta_y|^2+O(\Delta^6),
\end{align}
where $\alpha=4\alpha'$, $\beta=4|I_1+I_2|$, $\tilde\beta=16|I_3|$, and $\bar\beta=8|I_4|$. This is Eq.\ (3) of the main text.

As we said in the main text,  ${\bar \beta} > {\tilde \beta}$ and the action is unconstrained at the quartic level. We assume that  sixth order terms, given by
convolutions of six Green's functions have positive a prefactor $\gamma$ and constrain the action. An order of magnitude evaluation of  $\gamma$ gave
 $\gamma\sim 1/(v_F^2T^3)\sim1/(v_F^2g^3)$.

\subsection{B.~~~Stabilization of the mixed CDW/PDW order}

 In an $\su$-symmetric model, all states that satisfy $|\Delta_x|=|\Delta_y|=|\Delta|$ and $\Gamma=\Tr(U_AU_C^\dagger U_BU_D^\dagger)=-2$ are degenerate ground states.
 These include pure checkerboard CDW and PDW states and a mixed CDW/PDW state with stripe CDW and PDW orders between pairs of hot spots along $x$ and $y$ directions respectively (or vise versa).   We show that, when FS curvature is non-zero,  the mixed CDW/PDW order generates a secondary  homogeneous SC order, and this favors the mixed CDW/PDW order over pure CDW or PDW states.   We also show that the FS curvature lifts the degeneracy between CDW and PDW orders and makes both quadratic and quartic terms in the effective action anisotropic.  We show that this also favors the  mixed ``stripe" CDW/PDW state for some range of parameters.

\subsubsection{1.~~~Coupling to secondary homogeneous SC order}\label{second}
As we said in the main text, in the mixed state,
 CDW and PDW orders which carry the {\it same} momentum generically induce a secondary homogeneous SC order via triple coupling terms, and these terms can lower the Free energy. We show
 that the mixed CDW/PDW optimizes such coupling.
 The induced secondary SC order was shown in Ref.~\onlinecite{sm_pdw} to be a mixture of $s$-wave and $d_{xy}$-wave. We define $s$-wave SC and $d_{xy}$-wave SC order parameters as $\Psi_s$ and $\Psi_{d_{xy}}$. The action for the secondary SC order is given by~\cite{sm_pdw}
\begin{align}
\mathcal{S}_{\rm tri}[\Psi_s,\Psi_{d_{xy}},\rho,\psi]=&\alpha_s'(|\Psi_s|^2+|\Psi_{d_{xy}}|^2)-Y[(\Psi_s+\Psi_{d_{xy}})(\rho_A\psi_D^*+\rho_D^*\psi_A^*+ \rho_B^*\psi_C^*+\rho_C\psi_B^*)\nonumber\\
&+(\Psi_s-\Psi_{d_{xy}})( \rho_B\psi_D^*+\rho_D\psi_B^*+\rho_A^*\psi_C^*+\rho_C^*\psi_A^*)]+h.c.\nonumber\\
=&\frac{\alpha_s'}2(|\Psi_s+\Psi_{d_{xy}}|^2+|\Psi_s-\Psi_{d_{xy}}|^2)-Y[(\Psi_s+\Psi_{d_{xy}})(\rho_A\psi_D^*+\rho_D^*\psi_A^*+ \rho_B^*\psi_C^*+\rho_C\psi_B^*)\nonumber\\
&+(\Psi_s-\Psi_{d_{xy}})( \rho_B\psi_D^*+\rho_D\psi_B^*+\rho_A^*\psi_C^*+\rho_C^*\psi_A^*)]+h.c.,
\label{triple}
\end{align}
where $(\alpha'_s)^{-1}>0$ is the susceptibility of the secondary SC orders
 (for simplicity, we take this susceptibility to be the same for $\Psi_s$ and $\Psi_{d_{xy}}$, a qualitatively similar mixed CDW/PDW ground will still result if the two susceptibilities are not the same).
When both $\rho$ and $\psi$ are nonzero, superconducting orders $\Psi_s$ and $\Psi_{d_{xy}}$ are induced and the Free energy is lowered.

To minimize the Free energy (\ref{triple}), we maximize the magnitude of the two combinations of CDW/PDW order parameters $\rho_A\psi_D^*+\rho_D^*\psi_A^*+ \rho_B^*\psi_C^*+\rho_C\psi_B^*$ and $\rho_B\psi_D^*+\rho_D\psi_B^*+\rho_A^*\psi_C^*+\rho_C^*\psi_A^*$. We define
\begin{align}
\rho_A=|\Delta|\cos\theta_A e^{i\phi_A},~~&\psi_A=|\Delta|\sin\theta_A e^{i\bar\phi_A},\nonumber\\
\rho_B=|\Delta|\cos\theta_B e^{i\phi_B},~~&\psi_B=|\Delta|\sin\theta_B e^{i\bar\phi_B },\nonumber\\
\rho_C=|\Delta|\cos\theta_C e^{i\phi_C},~~&\psi_C=|\Delta|\sin\theta_C e^{i\bar\phi_C},\nonumber\\
\rho_D=|\Delta|\cos\theta_D e^{i\phi_D},~~&\psi_D=|\Delta|\sin\theta_D e^{i\bar\phi_D},
\end{align}
where $0<\theta_{A,B,C,D}<\pi/2$, $-\pi<\phi_{A,B,C,D}<\pi$, and $-\pi<\bar\phi_{A,B,C,D}<\pi$.

The first CDW/PDW combination term in Eq.\ (\ref{triple}) hence becomes
\begin{align}
&\rho_A\psi_D^*+\rho_D^*\psi_A^*+ \rho_B^*\psi_C^*+\rho_C\psi_B^*\nonumber\\
&=|\Delta|^2\[\cos\theta_A\sin\theta_De^{i(\phi_A-\bar\phi_D)}+\cos\theta_D\sin\theta_Ae^{-i(\phi_D+\bar\phi_A)}+\cos\theta_B\sin\theta_Ce^{-i(\phi_B+\bar\phi_C)}+\cos\theta_C\sin\theta_Be^{i(\phi_C-\bar\phi_B)}\]\nonumber\\
&=|\Delta|^2\[\cos\theta_A\sin\theta_De^{i\phi_1}+\cos\theta_D\sin\theta_Ae^{i\phi_2}+\cos\theta_B\sin\theta_Ce^{i\phi_3}+\cos\theta_C\sin\theta_Be^{i\phi_4}\]\nonumber\\
&=|\Delta|^2\left\{e^{i(\phi_1-\phi_2)/2}\[\sin(\theta_A+\theta_D)\cos\(\frac{\phi_1+\phi_2}{2}\)-i\sin(\theta_A-\theta_D)\sin\(\frac{\phi_1+\phi_2}{2}\)\]\right.\nonumber\\
&~~~~~~~~~~\left.+e^{-i(\phi_3-\phi_4)/2}\[\sin(\theta_C+\theta_B)\cos\(\frac{\phi_3+\phi_4}{2}\)-i\sin(\theta_C-\theta_B)\sin\(\frac{\phi_3+\phi_4}{2}\)\]\right\}.
\label{com}
\end{align}
where in the third line we have defined $\phi_1=\phi_A-\bar\phi_D$, $\phi_2=\phi_D+\bar\phi_A$, $\phi_3=\phi_B+\bar\phi_C$, and $\phi_4=\phi_C-\bar\phi_B$.

For generic $\phi_{1,2,3,4}$ the magnitude of Eq. (\ref{com}) is maximized  when
\begin{align}
\theta_A=0&,~~\theta_D=\frac{\pi}{2}{,~~\rm{or}~~} A\rightarrow D\nonumber\\
\theta_B=0&,~~\theta_C=\frac{\pi}{2}{,~~\rm{or}~~} B\rightarrow C.
\end{align}
Repeating the same arguments for the second combination term $\rho_B\psi_D^*+\rho_D\psi_B^*+\rho_A^*\psi_C^*+\rho_C^*\psi_A^*$ we obtain one more set of conditions
\begin{align}
\theta_B=0&,~~\theta_D=\frac{\pi}{2}{,~~\rm{or}~~} A\rightarrow D\nonumber\\
\theta_A=0&,~~\theta_C=\frac{\pi}{2}{,~~\rm{or}~~} B\rightarrow C.
\end{align}
Combining, we obtain
\begin{align}
\theta_A=\theta_B=0&,~~\theta_C=\theta_D=\frac{\pi}{2}{,~~\rm{or}~~} A,B\rightarrow C,D.
\label{theta}
\end{align}

Then $\rho_A\psi_D^*+\rho_D^*\psi_A^*+ \rho_B^*\psi_C^*+\rho_C\psi_B^*=|\Delta|^2\[e^{i(\phi_A-\bar\phi_D)}+e^{-i(\phi_B-\bar\phi_C)}\]$ and $\rho_B\psi_D^*+\rho_D\psi_B^*+\rho_A^*\psi_C^*+\rho_C^*\psi_A^*=|\Delta|^2\[e^{i(\phi_B-\bar\phi_D)}+e^{-i(\phi_A+\bar\phi_C)}\]$.
 Maximizing the magnitude of both these terms,  we obtain the conditions on $\phi$'s and $\bar\phi$'s as
\begin{align}
\phi_A+\phi_B+\bar\phi_C-\bar\phi_D=0.
\label{phi}
\end{align}
It is easy to verify that Eqs.\ (\ref{theta}) and (\ref{phi}) yields $|\rho_A|=|\rho_B|=|\psi_C|=|\psi_D|=|\Delta|$, $\rho_A\rho_B\psi_C\psi_D^*=|\rho_A\rho_B\psi_C\psi_D|$, or $|\rho_C|=|\rho_D|=|\psi_A|=|\psi_B|=|\Delta|$, $\rho_A\rho_B\psi_C\psi_D^*=|\rho_A\rho_B\psi_C\psi_D|$. These states are exactly the mixed CDW/PDW state we described in the main text, related by a lattice $C_4$ rotation.

Before we proceed, we remind that $\theta$'s, $\phi$'s, and $\bar\phi$'s are {\it not} free parameters -- they are constrained by the condition $\Gamma=\Tr(U_AU_C^\dagger U_BU_D^\dagger)=-2$, and Eqs.\ (\ref{theta}) and (\ref{phi}) have to be consistent with this condition. We recall that in our notations
\begin{align}
U_A=\(\begin{array}{cc}
\sin\theta_Ae^{i\bar\phi_A}&\cos\theta_Ae^{-i\phi_A}\\
-\cos\theta_Ae^{i\phi_A}&\sin\theta_Ae^{-i\bar\phi_A}
\end{array}\),&~~
U_B=\(\begin{array}{cc}
\sin\theta_Be^{i\bar\phi_B}&\cos\theta_Be^{i\phi_B}\\
-\cos\theta_Be^{-i\phi_B}&\sin\theta_Be^{-i\bar\phi_B}
\end{array}\),\nonumber\\
U_C=\(\begin{array}{cc}
\sin\theta_Ce^{i\bar\phi_C}&\cos\theta_Ce^{-i\phi_C}\\
-\cos\theta_Ce^{i\phi_C}&\sin\theta_Ce^{-i\bar\phi_C}
\end{array}\),&~~
U_D=\(\begin{array}{cc}
\sin\theta_De^{i\bar\phi_D}&\cos\theta_De^{i\phi_D}\\
-\cos\theta_De^{-i\phi_D}&\sin\theta_De^{-i\bar\phi_D}
\end{array}\).
\label{14}
\end{align}
Plugging Eqs.\ (\ref{theta}) and (\ref{phi}) into Eq.\ (\ref{14}) we find that indeed $\Gamma=\Tr(U_AU_C^\dagger U_BU_D^\dagger)=-2$. Hence, the mixed CDW/PDW state is truly a ground state.

So far for the minimization with respect to $\theta$'s we have assumed that $\phi_{1,2,3,4}$ are completely generic. For the special case when $\phi_1=-\phi_2$ and $\phi_3=-\phi_4$, the condition on $\theta$'s are less strict -- from Eq.\ (\ref{com}) we see that  in this case one only needs to satisfy
\begin{align}
\theta_A=\pi/2-\theta_D,~~&{\rm and}~~\theta_B=\pi/2-\theta_C\equiv\theta.\nonumber\\
\phi_1=-\phi_2&=\phi_3=-\phi_4.
\label{mon}
\end{align}
Doing the same for  $\rho_B\psi_D^*+\rho_D\psi_B^*+\rho_A^*\psi_C^*+\rho_C^*\psi_A^*$ term and combining with (\ref{mon}) we find that
\begin{gather}
\theta_A=\theta_B=\pi/2-\theta_C=\pi/2-\theta_D\equiv\theta,\label{thetaa}\\
\phi_A-\bar\phi_D=-\phi_D-\bar\phi_A=-\phi_B-\bar\phi_C=\phi_C-\bar\phi_B,\nonumber\\
\phi_B-\bar\phi_D=\phi_D-\bar\phi_B=-\phi_A-\bar\phi_C=-\phi_C-\bar\phi_A.
\label{16}
\end{gather}

The configuration in Eqs.\ (\ref{thetaa}) and (\ref{16}) would give the same ground state energy as our mixed CDW/PDW state and hence has to be considered. However, again one needs to check if Eq.\ (\ref{16}) is consistent with $\Gamma=-2$, which in terms of $\theta$,  $\phi$'s, and $\bar\phi$'s becomes
\begin{align}
\Ree&\[\frac14\sin^22\theta\(e^{i(\bar\phi_C-\bar\phi_A)}+e^{i(\phi_C-\phi_A)}\)\(e^{i(\bar\phi_D-\bar\phi_B)}+e^{i(\phi_B-\phi_D)}\)\right.\nonumber\\
&\left.-\(\sin^2\theta e^{-i(\bar\phi_A+\phi_C)}-\cos^2\theta e^{-i(\phi_A+\bar\phi_C)}\)\(\sin^2\theta e^{i(\bar\phi_B-\phi_D)}-\cos^2\theta e^{i(\bar\phi_D-\phi_B)}\)\] =-1.
\end{align}

For a generic $\theta$, this condition is satisfied if
\begin{align}
\bar\phi_A+\phi_C-\phi_A-\bar\phi_C=&\pi,\nonumber\\
\bar\phi_B-\phi_D-\bar\phi_D+\phi_B=&\pi,\nonumber\\
\bar\phi_A+\phi_C-\bar\phi_B+\phi_D=&0.
\label{17}
\end{align}
or
\begin{align}
\bar\phi_A+\phi_C-\phi_A-\bar\phi_C=&0,\nonumber\\
\bar\phi_B-\phi_D-\bar\phi_D+\phi_B=&0.\nonumber\\
\bar\phi_A+\phi_C-\bar\phi_B+\phi_D=&0.\nonumber\\
\bar\phi_C-\bar\phi_A=&\pi,\nonumber\\
\bar\phi_D-\bar\phi_B=&0.
\label{18}
\end{align}

We have verified that neither Eq.\ (\ref{17}) nor (\ref{18}) is consistent with Eq.\ (\ref{16}). Hence Eqs.\ (\ref{thetaa}) and (\ref{16}) do not correspond to a true ground state of the CDW/PDW order, and only Eqs.\ (\ref{theta}) and (\ref{phi}) are the correct conditions that minimize the Free energy in Eq.\ (\ref{triple}). Minimization of Eq.\ (\ref{triple}) with respect to $\Psi_s$ and $\Psi_{d_{xy}}$ yields $|\Psi_s|=|\Psi_{d_{xy}}|\sim Y|\Delta|^2/\alpha_s'$, and ${\mathcal S}_{\rm tri}=-\delta\beta|\Delta|^4$, where $\delta\beta\sim Y^2/\alpha_s'$. We can then effectively write down the Free energy (\ref{triple}) as
\begin{align}
S_{\rm tri}=-\delta\beta|\Delta|^4(\cos\theta_A\cos\theta_B\sin\theta_C\sin\theta_D+\sin\theta_A\sin\theta_B\cos\theta_C\cos\theta_D).
\end{align}

\subsubsection{2.~~~Other effects beyond hot spot approximation}\label{comb}

 Two other effects beyond hot spot approximation have been considered in Ref.~\onlinecite{sm_pdw}. First, at the quadratic level, there appears the term $\sim -\delta\alpha|\psi|^2$ in the action which favors PDW order.  Second, at the quartic level, coupling to fermions away from hot spots  reduces the CDW Free energy by $\sim-\delta\gamma|\rho_A|^2|\rho_B|^2$ and favors CDW order. These  two effects and the one considered in the previous subsection give rise to an extra piece in the effective action of the form
  \begin{align}
\delta \mathcal{S}=&-\delta\alpha|\Delta|^2(\sin^2\theta_A+\sin^2\theta_B+\sin^2\theta_C+\sin^2\theta_D)\nonumber\\
&-\delta\beta|\Delta|^4(\cos\theta_A\cos\theta_B\sin\theta_C\sin\theta_D+\sin\theta_A\sin\theta_B\cos\theta_C\cos\theta_D)\nonumber\\
&-\delta\gamma|\Delta|^4(\cos^2\theta_A\cos^2\theta_B+\cos^2\theta_C\cos^2\theta_D),
\label{1233}
\end{align}
where $\delta\alpha, \delta\beta,\delta\gamma>0$.
In the ground state we have $\theta_C=\theta_D\equiv \theta_x$ and $\theta_A=\theta_B\equiv \theta_y$. The values of $\theta_x$ and $\theta_y$ depend on the parameters $\delta\alpha$, $\delta\beta$, and $\delta\gamma$. If $\delta\alpha$ term is the largest, the ground state has $\theta_x=\theta_y=\pi/2$ and is a pure PDW state; if $\delta\gamma$ term is the largest one,  $\theta_x=\theta_y=0$ and the ground state has pure CDW. If $\delta\beta$ is the largest,  $\theta_{x,y}=0$, $\theta_{y,x}=\pi/2$ and the ground state is our mixed CDW/PDW. For this case it is easy to verify that the condition on $\delta\beta$
to stabilize a mixed CDW/PDW state is
\begin{align}
\delta\beta>\bigg|\delta\gamma-\frac{\delta\alpha}{|\Delta|^2}\bigg|.
\label{1234}
\end{align}
 We note that when this condition is satisfied, the ground state will no longer have equal magnitudes of the CDW and PDW orders (however, the ground state will remain a mixed CDW/PDW state). It is relevant for this reasoning that the transition into the mixed CDW/PDW state is first-order because if it was second-order,
 the condition (\ref{1234}) could not be satisfied at temperatures right below $T_{\rm cdw}$ and the system would  first develop a pure PDW order.

\subsection{C.~~~SC/BO sector}
In this subsection we analyze the interplay between $d_{x^2-y^2}$-wave SC and bond charge orders (BO) with momenta $(Q,\pm Q)$.
We define SC order parameters $\Psi_a\sim\langle c_1 c_6\rangle$, $\Psi_b\sim\langle c_2c_5\rangle$, and BO parameters $\Phi_a\sim\langle c_1^\dagger c_6\rangle$, $\Phi_b\sim\langle c_2^\dagger c_5\rangle$. All order parameters change sign when fermionic momenta are changed by $(\pi,\pi)$, namely, $\langle c_1 c_6\rangle-\langle c_3 c_8\rangle$, $\langle c_2c_5\rangle-\langle c_4c_7\rangle$, $\langle c_1^\dagger c_6\rangle-\langle c_3^\dagger c_8\rangle$, and $\langle c_2^\dagger c_5\rangle-\langle c_4^\dagger c_7\rangle$.
For $\su$- symmetric model the effective action can be written as~\cite{sm_efetov_1,23}
\begin{align}
\mathcal{S}_{\rm sc/bo}=&\alpha_s\(|\Psi_a|^2+|\Phi_a|^2\)+\alpha_s\(|\Psi_b|^2+|\Phi_b|^2\)+\beta_s\(|\Psi_a|^2+|\Phi_a|^2\)^2+\beta_s\(|\Psi_b|^2+|\Phi_b|^2\)^2.
\end{align}
This action has $\rm O(4)\times O(4)$ symmetry.
 The two SC order parameters  $\Psi_a$ and $\Psi_b$ can be combined into $d_{x^2-y^2}$ and $B_{2g}$ order parameters as
\begin{align}
\Psi_d&=(\Psi_a+\Psi_b)/\sqrt{2}{~~{\rm and}~~}\Psi_{B_{2g}}=(\Psi_a-\Psi_b)/\sqrt{2}.
\end{align}
Equivalently we have $\Psi_a=(\Psi_d+\Psi_{B_{2g}})/\sqrt{2}$, $\Psi_b=(\Psi_d-\Psi_{B_{2g}})/\sqrt{2}$.

 By obvious practical reasons, we only consider $d_{x^2-y^2}$ wave SC and set $\Psi_{B_{2g}}$ to zero.
  We then have $\Psi_1=\Psi_d/\sqrt{2}$, $\Psi_2=\Psi_d/\sqrt{2}$. The action takes the form
\begin{align}
\mathcal{S}_{\rm sc/bo}=&\alpha_s(|\Psi_d|^2+|\Phi_a|^2+|\Phi_b|^2)+\beta_s\[\(\frac{{|\Psi_d|}^2}2+|\Phi_a|^2\)^2+\(\frac{{|\Psi_d|}^2}2+|\Phi_b|^2\)^2\]\nonumber\\
\equiv&\alpha_s(|\Psi_d|^2+|\Phi_a|^2+|\Phi_b|^2)+\frac{\beta_s}{2}\(|\Psi_d|^2+|\Phi_a|^2+|\Phi_b|^2\)^2+\frac{\beta_s}{2}\(|\Phi_a|^2-|\Phi_b|^2\)^2.
\label{scbo}
\end{align}

The fist two terms in the last line of  Eq.\ (\ref{scbo}) describe a model with $\rm {O}(6)$ symmetry~\cite{sm_o6}. The last term breaks
 the $\rm {O}(6)$ symmetry and  gaps out would be longitudinal Goldstone mode between $|\Phi_a|$ and $|\Phi_b|$.
  As a result, the model of Eq.\ (\ref{scbo}) has four Goldstone modes instead of five for an $\rm O(6)$-symmetric model.
 Because $\beta_s>0$, for the ground state of Eq.\ (\ref{scbo}) we have $|\Phi_a|=|\Phi_b|$. Defining $\Phi\equiv\sqrt2\Phi_a$ and setting $|\Phi_a|=|\Phi_b|$, we obtain from (\ref{scbo})
\begin{align}
\mathcal{S}_{\rm sc/bo}[\Psi_d, \Phi]=&\alpha_s(|\Psi_d|^2+|\Phi|^2)+\frac{\beta_s}{2}\(|\Psi_d|^2+|\Phi|^2\)^2.
\label{scbo1}
\end{align}
In terms of $\Psi_d$ and $\Phi$, the ground state has an $\rm O(4)$ degeneracy.

The degeneracy is lifted once one includes into consideration the FS curvature at hot spots.
Then the action becomes
\begin{align}
\mathcal{S}_{\rm sc/bo}[\Psi_d, \Phi]=&\bar\alpha_s|\Psi_d|^2+\tilde\alpha_s|\Phi|^2+\frac{\beta_s}{2}\(|\Psi_d|^4+|\Phi|^2\)^2,
\end{align}
where $\bar\alpha_s<\tilde\alpha_s$. In this case the pure SC state minimizes the Free energy and becomes the true ground state of the SC/BO sector~\cite{sm_efetov_1,ms}.
BO always comes second and, because this order does not additionally break any discrete symmetry, there is no possibility to lift the instability temperature
 for this order above the superconducting $T_c$.

\subsection{D.~~~The coupling between $d_{x^2-y^2}$-wave SC and the mixed CDW/PDW order}\label{int}

In this subsection, we discuss the interplay between $d_{x^2-y^2}$-wave SC and the mixed CDW/PDW order.
 We first show that at low temperatures, but in the range where GL expansion is  applicable, the CDW/PDW gap $\Delta$ is parametrically larger than the SC gap $\Psi$.
  To see this, we note that the GL action for mixed CDW/PDW is $\mathcal{S}_{\rm cdw/pdw}=\alpha |\Delta|^2-\beta|\Delta|^4+\gamma|\Delta|^6$. Minimizing this action we obtain $|\Delta|^2=(\beta+\sqrt{\beta^2-3\gamma\alpha})/3\gamma$. Below $T_{\rm cdw}$,  $\alpha$ is negative and scales as $-\Lambda/v_F^2$. Then $\Delta^2\approx \sqrt{-\alpha/(3\gamma)}\sim g^2\sqrt{\Lambda/g}$. Meanwhile, the SC gap is given by $|\Psi|^2=|\alpha_s|/(2\beta_s)$, where $|\alpha_s|\sim \Lambda/v_F^2$ and $\beta_s\sim\Lambda/(v_FT_c)^2\sim \Lambda/(v_Fg)^2$. Using the fact that $T_c$ and $T_{\rm cdw}$ are both of order $g$, where, we recall,  $g$ is spin-fermion coupling,
  we obtain  $|\Psi|^2 \sim |\Delta^2|\sqrt{g/\Lambda}\ll |\Delta|^2$.
This result implies that
 the interaction with SC order does not distort substantially the inner structure of the CDW/PDW order. In particular, the condition
  $\Gamma=-2$ yields the minimum of the Free energy also in the presence of $d_{x^2-y^2}$ SC.

We now derive the effective action. We first note that $d_{x^2-y^2}$ SC and CDW/PDW orders do not couple via triple couplings $\sim \Psi\rho\psi$, because all three order parameters change sign when fermionic momenta are shifted by $(\pi, \pi)$, and the contributions from fermionic momenta that differ by $(\pi,\pi)$ cancel out.  Rather, $d_{x^2-y^2}$ SC couples separately to CDW and PDW components. We will drop the subscript $d$ and use $\Psi$ for $d_{x^2-y^2}$ SC order parameter.

\begin{figure}
\includegraphics[width=0.8\columnwidth]{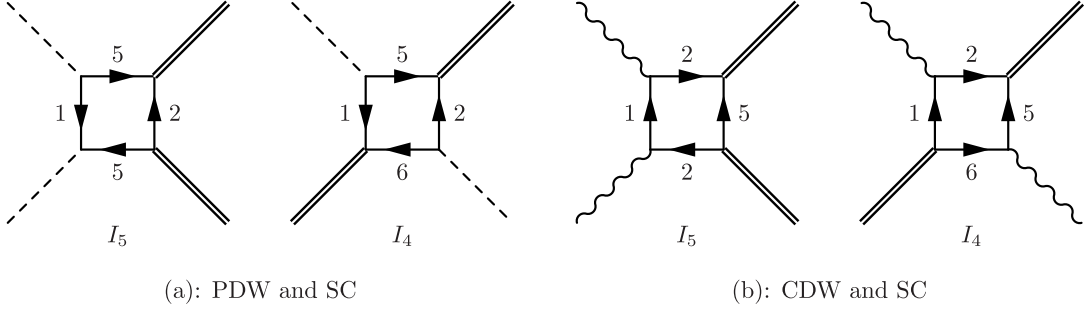}
\caption{Panel (a): the diagrams for coupling between PDW and SC orders. Panel (b): the diagrams for coupling between CDW and SC orders. We have omitted many diagrams of the same type. Dashed lines stand for PDW order, wavy lines stand for CDW order, and double lines stand for SC order.}
\label{i5}
\end{figure}

For PDW components, we find two types of couplings to SC, which are given by the diagrams shown in Fig.\ \ref{i5}(a). The first diagram yields a coupling term $-I_5|\psi_C|^2|\Psi|^2$, where
\begin{align}
I_5=&-T\sum_{\omega_m}\int \frac{d^2k}{4\pi^2} G_1(\omega_m, k)G_2(\omega_m, k)G_5^2(-\omega_m, -k).
\end{align}
Evaluating the integral in the same way as with did for $I_1-I_4$ earlier, we obtain~\cite{sm_23}
\begin{align}
I_5=&-T\sum_{\omega_m}\int\frac{d^2k}{4\pi^2}\frac{1}{i\omega_m-\epsilon_1(k)}\frac{1}{-i\omega_m-\epsilon_2(k)}\frac{1}{[-i\omega_m-\epsilon_5(-k)]^2}\nonumber\\
=&T\sum_{\omega_m}\int\frac{d\epsilon_1d\epsilon_2}{8\pi^2v_xv_y}\frac{1}{(\epsilon_1-i\omega_m)(\epsilon_2-i\omega_m)(\epsilon_2^2+\omega_m^2)}\nonumber\\
=&-T\sum_{\omega_m}\int\frac{d\epsilon_2}{8\pi v_xv_y}\frac{|\omega_m|}{(\epsilon_2^2+\omega_m^2)^2}\nonumber\\
=&-\frac{T}{16v_xv_y}\sum_{\omega_m}\frac{1}{\omega_m^2}=-\frac{1}{64v_xv_y}\frac{1}{T} = \frac{1}{2} I_4.
\end{align}
The second diagram in of Fig.\ \ref{i5}(a) yields the coupling term that depends on the phases of $\psi$ and $\Psi$. This term is
  $-I_4\psi_C\psi_D(\Psi^*)^2$.

Combining contributions from the two types of  diagrams, we obtain
\begin{align}
\mathcal{S}_{\rm pdw/sc}=2|I_4|(|\psi_C|^2+|\psi_D|^2)|\Psi|^2+2|I_4|[\psi_C\psi_D(\Psi^*)^2+\psi_C^*\psi_D^*\Psi^2].
\end{align}
Keeping in mind that $|\psi_C|=|\psi_D|$, we find that $\mathcal{S}_{\rm pdw/sc}$ is minimized  when
 $\psi_C\psi_D(\Psi^*)^2=-|\psi_C\psi_D(\Psi^*)^2|$. At the minimum,  $\mathcal{S}_{\rm pdw/sc} =0$, i.e.,
  the PDW component of the mixed CDW/PDW order does not couple to SC order.

For CDW components, the coupling terms  are similar, as we show in Fig.\ \ref{i5}(b). Following the same steps we obtain
\begin{align}
\mathcal{S}_{\rm cdw/sc}=2|I_4|(|\rho_A|^2+|\rho_B|^2)|\Psi|^2+2|I_4|(\rho_A\rho_B^*+\rho_A^*\rho_B)|\Psi|^2.
\end{align}
In distinction to PDW case,  the relative phase of $\rho_A$ and $\rho_B$ is locked at $\pm\pi/2$ due to coupling  to fermions away from hot spots~\cite{sm_charge}.
 For such a phase difference,  the second term in $\mathcal{S}_{\rm cdw/sc}$ vanishes and the coupling between CDW component and SC reduces to
\begin{align}
\mathcal{S}_{\rm cdw/sc}=4|I_5|(|\rho_A|^2+|\rho_B|^2)|\Psi|^2.
\end{align}
Using $|\rho_A|=|\rho_B|=|\Delta|$, we obtain $\mathcal{S}_{\rm int}=\mathcal{S}_{\rm pdw/sc}+\mathcal{S}_{\rm cdw/sc}=\beta'|\Delta|^2|\Psi|^2$ and $\beta'\sim |I_4|\sim 1/(v_F^2T)\sim1/(v_F^2g)$. This is the main result of this analysis.

\section{II~~~Details of the fermionic spectra}
In this section we present the details of our analysis of the feedback from the CDW/PDW order on the fermionic dispersion in the antinodal region.

\subsection{A.~~~CDW and PDW orders away from hot spots}
\begin{figure}
\includegraphics[width=0.4\columnwidth]{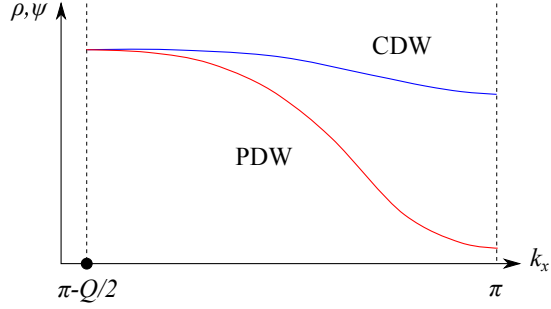}
\caption{Schematic plot of the CDW and PDW magnitudes as a function of $k_x$ of momentum in the antinodal region around momentum $(\pi,0)$,
obtained from Eqs.\ (\ref{pir}) and (\ref{pip}) at low temperatures $T\ll \epsilon_{1,2}$. Note that $k_x=\pi-Q/2$ is at the hot spot and $k_x=\pi$ is at the BZ boundary.}
\label{AN}
\end{figure}

 To calculate the fermionic dispersion in the whole antinodal region, we need to extend our analysis of the CDW and PDW orders to regions away from hot spots.
 We recall that in the $\su$-symmetric hot spot model, CDW and PDW components of the CDW/PDW order have equal magnitudes.
 Once we move away from hot spots, the equivalence gets lost.
The momentum dependence of CDW and PDW order parameters $\rho (k)=\delta_{\alpha\beta}\langle c_\alpha^\dagger({\bf Q}/2+ k) c_\beta(-{\bf Q}/2+k)\rangle$, and $\psi (k) = i\sigma_{\alpha\beta}^y\langle c_\alpha({\bf Q}/2+k) c_\beta({\bf Q}/2- k) \rangle$ at arbitrary $k$
 can, in principle,  be obtained by solving integral equations in momentum and frequency.
 However, solving such an integral equation  is a rather involved procedure.
 Here we adopt a simplified approach and just compare the kernels of the
 ladder equations for $\rho (k)$ and $\psi (k)$ on both sides of a hot spot, i.e., near the zone boundary and closer to zone diagonals.

Because in ladder series for $\rho (k)$ and $\psi (k)$ each spin-fermion interaction changes fermionic momenta by $\boldsymbol {\pi}=(\pi,\pi)$, we consider the effective kernel for small momentum transfer, made out of two subsequent terms in the ladder series.
For CDW order, such term is
\begin{align}
\Pi_\rho=
g^2\int_{\omega_m,k} G(\omega_m,{\bf Q}/2+{\boldsymbol {\pi}}+{\bf k}+\tilde k)G(\omega_m,-{\bf Q}/2+{\boldsymbol {\pi}}+{\bf k}+\tilde k)\int_{\omega_m',k'}G(\omega_m',{\bf Q}/2+{\bf k}+\tilde k')G(\omega_m',-{\bf Q}/2+{\bf k}+\tilde k'),
\end{align}
where ${\tilde k}$  is a deviation from a given ${\bf k}$. We assume that relevant ${\tilde k}$ are small enough and linearize fermionic dispersions in $\tilde k$.
For  PDW order, the product of two subsequent terms in the ladder series is
\begin{align}
\Pi_\psi= g^2 \int_{\omega_m,k} G(\omega_m,{\bf Q}/2+{\boldsymbol {\pi}}+{\bf k}+\tilde k)G(\omega_m,{\bf Q}/2+{\boldsymbol {\pi}}+{\bf k}-\tilde k)\int_{\omega_m',k'}G(\omega_m',{\bf Q}/2+{\bf k}+\tilde k')G(\omega_m',{\bf Q}/2+{\bf k}-\tilde k'),
\end{align}
 The effective interaction $g$ does indeed depend on momentum and also on frequency. However, when we consider fermions away from hot spots, this interaction
   is not singular and can be approximated by a constant.

We first consider ${\bf k}$ at the Brillouin zone boundary, i.e., at ${\bf k} = (\pi,0)$ if ${\bf Q} = (0, Q)$.
 We linearize the fermionic dispersion as $\epsilon({\bf Q}/2+{\boldsymbol {\pi}}+{\bf k}+\tilde k)=\epsilon_2+v_2k_y$, $\epsilon(-{\bf Q}/2+{\boldsymbol {\pi}}+{\bf k}+\tilde k)=\epsilon_2-v_2k_y$, $\epsilon({\bf Q}/2+{\bf k}+\tilde k)=\epsilon_1-v_1k_y$, and $\epsilon(-{\bf Q}/2+{\bf k}+\tilde k)=\epsilon_1+v_1k_y$. For antinodal fermions, we assume that $\epsilon_{1,2}\ll \Lambda$. Note that the Fermi velocities between both fermion pairs that differ in momentum by $\bf Q$ are antiparallel. Evaluating the integrals, we obtain at small $T$:
\begin{align}
\Pi_{\rho}=&g^2T\sum_m\int\frac{d^2k}{4\pi^2}\frac{1}{(i\omega_m-\epsilon_2-v_2k_y)(i\omega_m-\epsilon_2+v_2k_y)}T\sum_{m'}\int\frac{d^2k'}{4\pi^2}\frac{1}{(i\omega_m'-\epsilon_1+v_1k_y')(i\omega_m'-\epsilon_1-v_1k_y')}\nonumber\\
=&g^2\Lambda_k^2\sum_m\int\frac{dk_y}{4\pi^2}\frac{1}{v_2^2k_y^2+(\omega_m+i\epsilon_2)^2}\sum_{m'}\int\frac{dk_y'}{4\pi^2}\frac{1}{v_1^2(k_y')^2+(\omega_m'+i\epsilon_1)^2}=\frac{g^2\Lambda_k^2}{64\pi^4v_1v_2}\log\frac{\Lambda}{\epsilon_2}\log\frac{\Lambda}{\epsilon_1},
\label{pir}
\end{align}
and
\begin{align}
\Pi_{\psi}=&g^2T\sum_m\int\frac{d^2k}{4\pi^2}\frac{1}{(i\omega_m-\epsilon_2-v_2k_y)(-i\omega_m-\epsilon_2+v_2k_y)}T\sum_{m'}\int\frac{d^2k'}{4\pi^2}\frac{1}{(i\omega_m'-\epsilon_1+v_1k_y')(-i\omega_m'-\epsilon_1-v_1k_y')}\nonumber\\
=&g^2\Lambda_k^2\sum_m\int\frac{dk_y}{4\pi^2}\frac{1}{\epsilon_2^2+(\omega_m+iv_2k_y)^2}
\sum_{m'}\int\frac{dk_y'}{4\pi^2}\frac{1}{\epsilon_1^2+(\omega_m-iv_1k_y')^2}
\approx\begin{cases} 0 &\mbox{if } \Lambda\ll v_{1,2}\Lambda_k \\
\frac{g^2\Lambda_k^2}{16\pi^4v_1v_2} & \mbox{if } \Lambda\gg v_{1,2}\Lambda_k. \end{cases}
\label{pip}
\end{align}
Here $\Lambda$ and $\Lambda_k$ are  energy and momentum cutoffs, respectively.  We see that $\Pi_{\rho}\gg\Pi_{\psi}$.
 As the consequence, at the zone boundary the magnitude of the CDW order is much larger than for PDW order. This result can be straightforwardly understood because in the CDW channel Fermi velocities of the  two fermions which form the condensate are antiparallel, while in the PDW channel they are  parallel.

 As we move towards hot spots, the difference between the magnitudes of CDW and PDW orders gets weaker, and when ${\bf k}$  is between hot spots,
   CDW and PDW orders become degenerate.
    If we move further away from zone boundary, the kernels in PDW and CDW channels remain comparable to each other.
    Therefore, moving from the BZ boundary toward BZ center, the PDW gap should increase and finally become the same with CDW gap at the hot spot, as we sketch in Fig.\ \ref{AN}. This is the main conclusion of this Subsection.

\subsection{B.~~~Feedback from the mixed CDW/PDW order on the dispersion of antinodal fermions}
\begin{figure}
\includegraphics[width=0.9\columnwidth]{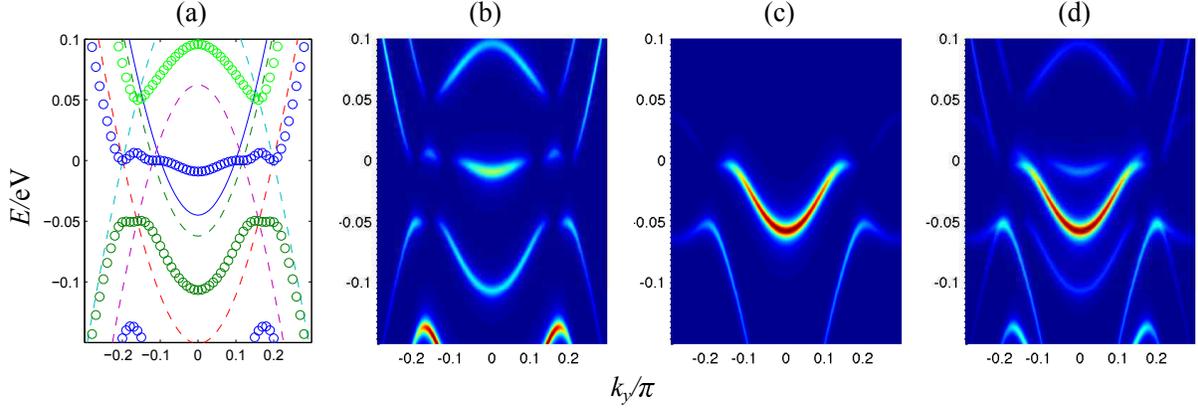}
\caption{Panel (a): The dispersions of energy eigenstates in the presence of CDW and PDW orders for Domain II at $k_x=0.9\pi$ and $\rho=\psi=\Delta=80~\rm meV$. In general $\rho$ and $\psi$ are not equal and depend on momentum, but this does not change our conclusion. Panel (b): The spectral function in the presence of CDW and PDW orders for Domain II for the same parameters. Near the energy of the original band, the shadow bands and the original band strongly hybridize and the spectral weight of the original band is hence reduced.  Panel (c): For comparison, the spectral function for Domain I at $k_x=0.9\pi$. Panel (d): The spectral function for Domain I and Domain II combined. The features from Domain I are clearly visible.}
\label{arpes}
\end{figure}

The feedback from a pure CDW order on the fermionic dispersion has been analyzed in Ref.~\onlinecite{sm_charge}. We follow the same approach detailed there, only in this work we also include PDW off-diagonal matrix elements. The  matrix from whose determinant one can extract  the fermionic energy  contains the original band  and the shadow bands with momentum shifted by  $\pm Q, \pm 2Q, ...$.
For our purposes it is sufficient to only retain four shadow bands: two from CDW coupling with momentum $(0,\pm Q)$, and two from PDW coupling with same momentum, because in the antinodal region all other shadow bands have high energy.
Therefore, the energy matrix to diagonalize becomes 5$\times$5 and we have 
\begin{align}
E(k)=\(\begin{array}{ccccc}
\epsilon(k)&\rho_Q(k)&\rho_{-Q}( k)&\psi_{Q}(k)&\psi_{-Q}(k)\\
\rho_Q^*(k)&\epsilon(k+Q)&0&0&0\\
\rho_{-Q}^*(k)&0&\epsilon(k-Q)&0&0\\
\psi_{Q}^*(k)&0&0&-\epsilon(-k+Q)&0\\
\psi_{-Q}^*(k)&0&0&0&-\epsilon(-k-Q)
\end{array}\),
\label{mat}
\end{align}
where $Q=(0,Q)$. Since we are mainly interested in the fermionic spectra close to the Fermi surface, which is almost ``horizontal" for the antinodal region close to $(\pi,0)$, we simply take $\rho$ and $\psi$ to be functions of $k_x$, i.e., $\rho_Q(k)=\rho_{-Q}( k)=\rho(k_x)$ and $\psi_Q(k)=\psi_{-Q}( k)=\psi(k_x)$.

  For the input values for CDW and PDW order parameters $\rho(k_x)$ and $\psi(k_x)$, we use the results from the previous subsection. We set CDW order parameter $\rho (k)$ to be independent on ${\bf k}$ in the antinodal region, and assume that PDW order parameter is zero at the Brillouin zone boundary and increases and approaches the CDW order as one moves the  scan  at fixed $k_x$ from the zone boundary towards the one which passes through a hot spot.  Specifically, we set
 \begin{align}
\rho(k_x)=\Delta{~~{\rm and}~~}
\psi(k_x)=\frac{2(\pi-k_x)}{Q}\Delta
\end{align}
We  take $\Delta=80~\rm meV$ to match the ARPES data.
As experimental input, we use the dispersion  in nearly optimally doped Pb-Bi2201 from Ref.~\onlinecite{sm_shen_a}: $\epsilon(k_x, k_y)=-2t(\cos{k_x}+\cos{k_y})-4t^{'} (\cos{k_x} \cos{k_y})-2 t^{''}
(\cos{2k_x}+\cos{2k_y})-4 t^{'''}(\cos{2k_x} \cos{k_y}+\cos{k_x} \cos{2k_y})-\epsilon_0$, with
$t =0.22 {\rm eV}$, $t^{'} = -0.034315 {\rm eV}$, $t^{''} = 0.035977 {\rm eV}$, $t^{'''} = -0.0071637 {\rm eV}$ , and $\epsilon_0 = -0.240577 {\rm eV}$. For this material, the CDW wave-vector is $Q=0.3\pi$ (see Ref.~\onlinecite{sm_hudson}). We diagonalized the 5 by 5 energy matrix (\ref{mat}) and computed the spectral function $I(\omega,k)\propto{\Imm}(\langle c_k (\omega) c^{\dagger}_{k} (\omega)\rangle)$. Our results are shown in Fig.\ 2 of the main text.

 Because the  CDW/PDW order breaks $C_4$ symmetry,  the measured spectra should be a combination of contributions from two domains~\cite{sm_charge}. So far we have only considered one domain (Domain I) with the ordering momentum ${\bf Q}=(0,Q)$. For the other domain  with ordering momentum ${\bf Q}=(Q,0)$  (Domain II) we found that for most of momentum range the CDW and PDW shadow bands and the original band have comparable energies [see Fig.\ \ref{arpes}(a)] in which case
  the spectrum  is ``dimmer" and features are much less pronounced. For comparison, in Fig.\ \ref{arpes} we present the spectral function from
  $k_x=0.9\pi$ separately  from the two domains and their combination. We see that the visible features are those from Domain I, i.e., the contribution from Domain II can be safely neglected.  We found similar results  for other values of $k_x$.

 \subsection{C.~~~Incompatibility of the a pure CDW state with the ARPES data}
 For comparison, we present the calculated fermionic spectra along 
 %YW same cuts in Fig.\ 2(b)
 multiple cuts like we did in Fig.\ 2(b)
  for a pure CDW state, and show that it is generally inconsistent with APPES data. Such an argument was first made by P. A. Lee~\cite{sm_patrick}, and we  reproduce it here.

\begin{figure}
 \includegraphics[width=0.68\columnwidth]{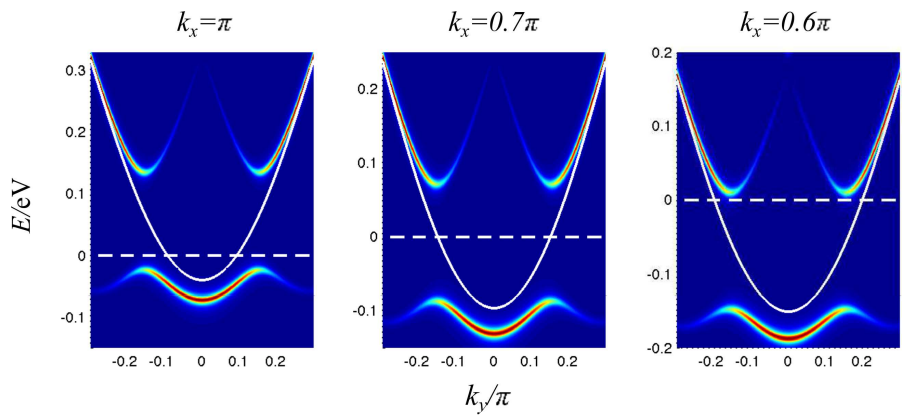}
 \caption{The computed fermionic spectral function for $k_x=\pi$, $k_x=0.7\pi$ and $k_x=0.6\pi$, assuming that only CDW order is present. The white parabolic line corresponds to the bare dispersion. Note that the dispersion crosses the Fermi level by a state moving from the upper branch, leaving a gap below the Fermi level.}
 \label{arp3}
 \end{figure}

 We perform the same calculation as in previous Subsection, the only difference being that the PDW order parameter $\psi$ has been set to zero from the very beginning. We show the resulting fermionic dispersions for $k_x=\pi$, $k_x=0.7\pi$ and $k_x=0.6\pi$ in Fig.\ \ref{arp3}.

 Comparing this with  Fig. 2(c) in the main text, we see that at $k_x=\pi$ our spectral function is generally consistent with experimental data. However, for other $k_x$ values, there is clear discrepancy. Namely, in the ARPES data, the Fermi arc region begins by ``the lower branch moving up", while in our result with purely CDW order, the Fermi arc begins by ``the upper branch moving down", leaving a finite gap below the Fermi level. This discrepancy is remedied by the presence of a PDW order, as we demonstrated in the previous subsection.


\begin{thebibliography}{99}
\bibitem{mark_last}
Tao Wu,	 Hadrien Mayaffre, Steffen Kr\"amer, Mladen Horvati\'c, Claude Berthier, W. N. Hardy, Ruixing Liang, D. A. Bonn, and Marc-Henri Julien, {Nature} {\bf 477}, 191-194 (2011);
T. Wu, H. Mayaffre, S. Kr\"amer, M. Horvati\'c, C. Berthier, W.N. Hardy, R. Liang, D.A. Bonn, and M.-H Julien, Nat. Comm. {\bf 6}, 6438 (2015).

\bibitem{ybco} G. Ghiringhelli, M. Le Tacon, M. Minola, S. Blanco-Canosa, C. Mazzoli, N.B. Brookes, G.M. De Luca, A. Frano, D. G. Hawthorn, F. He, T. Loew, M. Moretti Sala, D.C. Peets, M. Salluzzo, E. Schierle, R. Sutarto, G. A. Sawatzky, E. Weschke, B. Keimer, and L. Braicovich, {Science}, {\bf 337,  }821 (2012).

\bibitem{ybco_1} A. J. Achkar, R. Sutarto, X. Mao, F. He, A. Frano, S. Blanco-Canosa, M. Le Tacon, G. Ghiringhelli, L. Braicovich, M. Minola, M. Moretti Sala, C. Mazzoli, Ruixing Liang, D. A. Bonn, W. N. Hardy, B. Keimer, G. A. Sawatzky, and D. G. Hawthorn, Phys. Rev. Lett., {\bf 109}, 167001 (2012).

\bibitem{X-ray} R. Comin, A. Frano, M. M. Yee, Y. Yoshida, H. Eisaki, E. Schierle, E. Weschke, R. Sutarto, F. He, A. Soumyanarayanan, Y. He, M. Le Tacon, I. S. Elfimov, J. E. Hoffman, G. A. Sawatzky, B. Keimer, and A. Damascelli, Science {\bf 343}, 390-392 (2014).

\bibitem{X-ray_1} E. H. da Silva Neto, P. Aynajian, A. Frano, R. Comin, E. Schierle, E. Weschke, A. Gyenis, J. Wen, J. Schneeloch, Z. Xu, S. Ono, G. Gu, M. Le Tacon, A. Yazdani,  Science 343, 393-396 (2014).

\bibitem{davis_1}K. Fujita, M. H. Hamidian, S. D. Edkins, C. K. Kim, Y. Kohsaka, M. Azuma, M. Takano, H. Takagi, H. Eisaki, S. Uchida, A. Allais, M. J. Lawler, E.-A. Kim, S. Sachdev, and J. C. S\'eamus Davis,  PNAS {\bf 111} (30) E3026 (2014).

\bibitem{tranquada} J. M. Tranquada, G. D. Gu, M. H\"ucker, Q. Jie, H.-J. Kang, R. Klingeler, Q. Li, N. Tristan, J. S. Wen, G. Y. Xu, Z. J. Xu, J. Zhou, and M. v. Zimmermann, {Phys. Rev. B} {\bf  78,  }174529 (2008).

\bibitem{ber09} E. Berg, E. Fradkin, S.A. Kivelson, and J.M. Tranquada, New Journal of Physics {\bf 11}, 115004 (2009).

\bibitem{shen_a} Rui-Hua He, et al., Science {\bf 331}, 1579 (2011).

\bibitem{shen_2010}  M. Hashimoto,	Rui-Hua He,	K. Tanaka,	J.P. Testaud, W. Meevasana,	R. G. Moore, D. Lu,	H. Yao,	Y. Yoshida,	H. Eisaki, T. P. Devereaux,	Z. Hussain, and Z.X. Shen, Nature Physics {\bf 6}, 414 (2010).

\bibitem{kerr} J. Xia, E. Schemm, G. Deutscher, S. A. Kivelson, D. A. Bonn, W. N. Hardy, R. Liang, W. Siemons, G. Koster, M. M. Fejer, and A. Kapitulnik {Phys. Rev. Lett.} {\bf 100}, 127002 (2008); H. Karapetyan, J. Xia, M. H\"ucker, G. D. Gu, J. M. Tranquada, M. M. Fejer, and A. Kapitulnik, { Phys. Rev. Lett.} {\bf 112}, 047003 (2014).

\bibitem{bourges} Y. Sidis and P. Bourges, in {\it 10th International Conference on Materials and Mechanisms of Superconductivity (M2S-X), Washington DC, USA}, Journal of Physics: Conference Series Vol. 449 (IOP, Bristol, 2013);
 Yuan Li, V. Bal\'edent, G. Yu, N. Bari\v{s}i\'c, K. Hradil, R. A. Mole, Y. Sidis, P. Steffens, X. Zhao, P. Bourges, and M. Greven, {Nature} {\bf 468,  }283 (2010).

\bibitem{armitage} Y. Lubashevsky, LiDong Pan, T. Kirzhner, G. Koren, and N. P. Armitage, Phys. Rev. Lett. {\bf 112}, 147001, (2014).

\bibitem{ando} Y. Ando, K. Segawa, S. Komiya, and A. N. Lavrov  Phys. Rev. Lett. {\bf 88}, 137005 (2001).

\bibitem{hinkov} V. Hinkov, P. Bourges, S. Pailh\`es2, Y. Sidis, A. Ivanov, C. D. Frost, T. G. Perring, C. T. Lin, D. P. Chen, and B. Keimer, {Nat. Phys.} {\bf  3}, 780 (2007).

\bibitem{taillefer_last} O.\ Cyr-Choini\`ere, G.\ Grissonnanche, S.\ Dufour-Beaus\'ejour, S.\ Badoux, B.\ Michon, J. Day, D.\ A.\ Bonn, W. N. Hardy, R.\ Liang, N.\ Doiron-Leyraud, and L.\ Taillefer, {\it private communication}.

\bibitem{grilli} C. Castellani, C. Di Castro, and M. Grilli, Phys. Rev. Lett. {\bf 75}, 4650 (1995);  A. Perali, C. Castellani, C. Di Castro, and M. Grilli, Phys. Rev. B {\bf 54}, 16216 (1996).

\bibitem{ddw} S. Chakravarty, R. B. Laughlin, D. K. Morr, and C. Nayak. Phys. Rev. B 63, 094503 (2001).

\bibitem{ms} M. A. Metlitski and S. Sachdev, {  Phys. Rev. B} {\bf 82}, 075128 (2010).

\bibitem{efetov}  K. B. Efetov, H. Meier and C. P\'epin, {  Nat. Phys.} {\bf 9}, 442 (2013);  H. Meier, C. Pepin, M. Einenkel and K.B. Efetov, Phys. Rev. B {\bf 89}, 195115 (2014); K.B. Efetov {\it private communication}.

\bibitem{greco} A. Greco and M. Bejas, Phys. Rev. B {\bf 83}, 212503 (2011).

\bibitem{laplaca} S. Sachdev and R. L. Placa, Phys. Rev. Lett. {\bf 111}, 027202 (2013).

\bibitem{charge} Y. Wang and A. V. Chubukov, Phys. Rev. B {\bf 90}, 035149 (2014).

\bibitem{tsvelik}  A. M Tsvelik and A. V. Chubukov, Phys. Rev. B {\bf 89}, 184515 (2014).

\bibitem{norman} A. Melikyan and  M. R. Norman, Phys. Rev. B {\bf 89}, 024507 (2014).

\bibitem{debanjan}  D. Chowdhury and S. Sachdev, Phys. Rev. B {\bf 90}, 134516 (2014)

\bibitem{pepin} C. P\'epin, V. S. de Carvalho, T. Kloss, X. Montiel, Phys. Rev. B {\bf 90} 195207 (2014).

\bibitem{charge_1} Y. Wang, D. F. Agterberg,  and A. V. Chubukov, Phys. Rev. B {\bf 91}, 115103 (2015).

\bibitem{atkinson} S. Bulut, W. A. Atkinson, and A. P. Kampf, Phys. Rev. B {\bf 88}, 155132 (2013); W. A. Atkinson, A. P. Kampf, and S. Bulut, New Journal of Physics {\bf 17}, 013025 (2015)

\bibitem{rahul} Y. Wang, A. V. Chubukov,  and R. Nandkishore,  Phys. Rev. B 90, 205130 (2014).

\bibitem{debanjan_1} D. Chowdhury and S. Sachdev, Phys. Rev. B 90 (24), 245136 (2014).

\bibitem{pepin_new}T. Kloss, X. Montiel, C. Pépin, arXiv:1501.05324 (2015).

\bibitem{kivelson} E. Fradkin and  S. A. Kivelson, Nature Physics 8, 865-866 (2012); E. Fradkin, S. A. Kivelson, J. M. Tranquada, arXiv:1407.4480 (2014).

\bibitem{agterberg} D.F. Agterberg, D.S. Melchert, and M.K. Kashyap, Phys. Rev. B {\bf 91}, 0545021 (2015).

\bibitem{agterberg_2} D.F. Agterberg and H. Tsunetsugu, Nat. Phys. {\bf 4},  639 (2008).

\bibitem{patrick} P.\ A.\ Lee, Phys. Rev. X {\bf 4}, 031017 (2014).

\bibitem{lee_senthil}T. Senthil and P. A. Lee, Phys. Rev. Lett. {\bf 103}, 076402 (2009).

\bibitem{corboz} P. Corboz, T.M. Rice, and M. Troyer, Phys. Rev. Lett. {\bf 113}, 046402 (2014).

\bibitem{varma} Z. Wang, G. Kotliar, and X.-F. Wang, {  Phys. Rev. B} {\bf  42}, 8690 (1990); C. M. Varma, {  Phys. Rev. B} 55 14554 (1997).

\bibitem{sudip} S. Chakravarty, R. B. Laughlin, D. K.  Morr, and C. Nayak, {  Phys. Rev. B} {\bf  63}, 094503 (2001).

\bibitem{proust} N. Doiron-Leyraud,	S. Badoux,	S. René de Cotret,	S. Lepault,	D. LeBoeuf,	F. Laliberté,	E. Hassinger,	B. J. Ramshaw,	D. A. Bonn,	W. N. Hardy,	R. Liang,	J.-H.. Park,	D. Vignolles,	B. Vignolle,	L. Taillefer, and C. Proust, Nat. Comm. {\bf 6}, 6032 (2015).

\bibitem{allias} A. Allais, J. Bauer and S. Sachdev, Phys. Rev. B {\bf 90}, 155114 (2014).

\bibitem{comm}  More accurately, the system either develops a checkerboard order or a stripe order which preserves time-reversal and mirror
 symmetries.  The first necessary develops in the hot spot model, the second may develop due to contributions from fermions away from hot regions~\cite{laplaca,charge}.

\bibitem{acs} Ar. Abanov, A. V. Chubukov, and M. A. Finkelstein, {  Europhys. Lett.} {\bf 54}, 488 (2001);
Ar. Abanov, A. V. Chubukov, and J. Schmalian, {Adv. Phys.} \textbf{52,} 119 (2003).

\bibitem{wang} Y. Wang and A. V. Chubukov, Phys. Rev. Lett. {\bf 110}, 127001 (2013).

\bibitem{wang_el} Y. Wang and A. V. Chubukov, Phys. Rev. B {\bf 88}, 024516 (2013).

\bibitem{sm} See Supplemental Material [url], which includes Refs.\ \onlinecite{23,o6,hudson}.

\bibitem {23} D. Chowdhury and S. Sachdev, Phys. Rev. B {\bf 90}, 134516 (2014).

\bibitem{o6} L. E. Hayward, D. G. Hawthorn, R. G. Melko, and S. Sachdev, Science {\bf 343}, 1336 (2014).

\bibitem{hudson}W.D. Wise, M.C. Boyer, K. Chatterjee, T. Kondo, T. Takeuchi, H. Ikuta, Y. Wang, E.W. Hudson, Nat. Phys. {\bf 4}, 696 (2008).

\bibitem{pnictides} Note in passing that a similar situation holds in doped Fe-pnictides, see
 A. E. B\"ohmer et al, arXiv:1412.7038;  J. Kang, X. Wang, A. V. Chubukov, and R. M. Fernandes, Phys. Rev. B {\bf 91}, 121104(R) (2015).

\bibitem{hinojosa} A. Hinojosa, R.M. Fernades, and A.V. Chubukov, Phys. Rev. Lett. {\bf 113}, 167001 (2014).

\bibitem{shen_b}  I. M. Vishik,
M. Hashimoto, R.-H. He, W.-S. Lee, F. Schmitt, D. Lu,
R. G. Moore, C. Zhang, W. Meevasana, T. Sasagawa,
et al., Proc. Nat. Acad. of Sci. {\bf 109}, 18332 (2012); E. Razzoli, G.
Drachuck, A. Keren, M. Radovic, N. C. Plumb, J. Chang,
Y.-B. Huang, H. Ding, J. Mesot, and M. Shi, Phys. Rev.
Lett. {\bf 110}, 047004 (2013); Y. Peng, J. Meng, D.Mou, J.
He, L. Zhao, Y.Wu, G. Liu, X. Dong, S. He, J. Zhang, et
al., Nat. Comm. {\bf 4}, 2459 (2013); Y-M. Lu, T. Xiang, and D.-H. Lee, Nat. Phys. {\bf 10}, 634 (2014).

\bibitem{sakai} S. Sakai, S. Blanc, M. Civelli, Y. Gallais, M. Cazayous, M.-A. Méasson, J. S. Wen, Z. J. Xu, G. D. Gu, G. Sangiovanni, Y. Motome, K. Held, A. Sacuto, A. Georges, and M. Imada,
Phys. Rev. Lett. {\bf 111}, 107001 (2013).

%AC added

%\bibitem{comm_2}  To get a non-zero magnitude of SC order one needs to include the curvature of the FS at hot spots (see Ref. \cite{charge_1}).

\end{thebibliography}

\begin{thebibliography}{99}

\bibitem{sm_pdw} Y. Wang, D. F. Agterberg, and A. V. Chubukov, Phys.
Rev. B {\bf 91}, 115103 (2015).
\bibitem {sm_23} D. Chowdhury and S. Sachdev, Phys. Rev. B {\bf 90}, 134516 (2014).
\bibitem{sm_charge} Y. Wang and A. Chubukov, Phys. Rev. B {\bf 90}, 035149 (2014).
\bibitem{sm_efetov_1}K. B. Efetov, H. Meier and C. P\'epin, Nat. Phys. {\bf 9}, 442 (2013).
\bibitem{sm_ms}M. A. Metlitski and S. Sachdev, Phys. Rev. B {\bf 82}, 075128 (2010).
\bibitem{sm_o6} L. E. Hayward, D. G. Hawthorn, R. G. Melko, and S. Sachdev, Science {\bf 343}, 1336 (2014).
\bibitem{sm_shen_a} Rui-Hua He, et al., Science 331, 1579 (2011).
\bibitem{sm_hudson}W.D. Wise, M.C. Boyer, K. Chatterjee, T. Kondo, T. Takeuchi,
H. Ikuta, Y. Wang, E.W. Hudson, Nat. Phys. {\bf 4}, 696 (2008).
\bibitem{sm_patrick} P. A. Lee, Phys. Rev. X, {\bf 4}, 031017 (2014).
\end{thebibliography}
\end{document}